\documentclass[fleqn,usenatbib,useAMS]{mnras}

\usepackage[T1]{fontenc}
\usepackage{ae,aecompl}
\usepackage{graphicx}	
\usepackage{amsmath}	
\usepackage{amssymb}	
\usepackage{xspace}
\usepackage[table]{xcolor}

\usepackage{scalerel}
\usepackage{tikz}
\usetikzlibrary{svg.path}
\definecolor{orcidlogocol}{HTML}{A6CE39}
\tikzset{
  orcidlogo/.pic={
    \fill[orcidlogocol] svg{M256,128c0,70.7-57.3,128-128,128C57.3,256,0,198.7,0,128C0,57.3,57.3,0,128,0C198.7,0,256,57.3,256,128z};
    \fill[white] svg{M86.3,186.2H70.9V79.1h15.4v48.4V186.2z}
                 svg{M108.9,79.1h41.6c39.6,0,57,28.3,57,53.6c0,27.5-21.5,53.6-56.8,53.6h-41.8V79.1z M124.3,172.4h24.5c34.9,0,42.9-26.5,42.9-39.7c0-21.5-13.7-39.7-43.7-39.7h-23.7V172.4z}
                 svg{M88.7,56.8c0,5.5-4.5,10.1-10.1,10.1c-5.6,0-10.1-4.6-10.1-10.1c0-5.6,4.5-10.1,10.1-10.1C84.2,46.7,88.7,51.3,88.7,56.8z};
  }
}

\usepackage{newtxtext,newtxmath}

\newcommand\orcidicon[1]{\href{https://orcid.org/#1}{\mbox{\scalerel*{
\begin{tikzpicture}[yscale=-1,transform shape]
\pic{orcidlogo};
\end{tikzpicture}
}{|}}}}

\newcommand{\vpeak}{\ensuremath{V_{\rm peak}}\xspace}

\newcommand{\Mpeak}{\ensuremath{M_{\rm peak}}\xspace}
\newcommand{\Mhalo}{\ensuremath{M_{\rm h}}\xspace}

\newcommand{\Msunh}{\ensuremath{h^{-1}\mathrm{M}_\odot}\xspace}

\newcommand{\Mpch}{\ensuremath{h^{-1}\mathrm{Mpc}}\xspace}
\newcommand{\hMpc}{\ensuremath{h^{-1}\mathrm{Mpc}}\xspace}

\definecolor{hpurple}{HTML}{7E16DF}

\newcommand{\logMh}{\ensuremath{\log_{10} (M_{\rm h}[h^{-1}\mathrm{M}_\odot])}\xspace}
\newcommand{\logMstar}{\ensuremath{\log_{10} (M_\star[h^{-1}\mathrm{M}_\odot])}\xspace}

\newcommand{\lcdm}{\ensuremath{\Lambda{\rm CDM}}\xspace}

\newcommand{\tng}{TNG\xspace}
\newcommand{\dtng}{Dark-TNG\xspace}
\newcommand{\bosstng}{BOSS-TNG\xspace}

\newcommand{\lil}{{\it lensing is low}\xspace}

\newcommand{\wps}{\ensuremath{\omega_\mathrm{p}}\xspace}
\newcommand{\ds}{\ensuremath{\Delta\Sigma}\xspace}

\definecolor{Gray}{gray}{0.9}
\newcolumntype{a}{>{\columncolor{Gray}}c}

\newcommand{\bit}{\begin{enumerate}}
\newcommand{\eit}{\end{enumerate}}

\defcitealias{lange2019_NewPerspectivesBOSS}{L19}
\defcitealias{amon2022ConsistentLensingClustering}{AR22}

\title[The origin of the lensing-is-low problem]{The galaxy formation origin of the \textbf{\textit{lensing is low}} problem}

\author[Chaves-Montero et al.]{
Jon\'{a}s Chaves-Montero$^{1,2}$\thanks{E-mail: \href{mailto:jchaves@ifae.es}{jchaves@ifae.es}}\orcidicon{0000-0002-9553-4261},
Raul E. Angulo$^{1}$\thanks{E-mail: \href{mailto:reangulo@dipc.org}{reangulo@dipc.org}} and
Sergio Contreras$^{1}.$
\\$^{1}$ Donostia International Physics Centre, Paseo Manuel de Lardizabal 4, 20018 Donostia-San Sebastian, Spain.
\\$^{2}$ Institut de F\'isica d'Altes Energies, The Barcelona Institute of Science and Technology, Campus UAB, E-08193 Bellaterra (Barcelona), Spain.
}

\date{Accepted XXX. Received YYY; in original form ZZZ}
\pubyear{2022}

\begin{document}
\label{firstpage}
\pagerange{\pageref{firstpage}--\pageref{lastpage}}
\maketitle

\begin{abstract}
Recent analyses show that $\Lambda$CDM-based models optimised to reproduce the clustering of massive galaxies overestimate their gravitational lensing by about 30\%, the so-called {\it lensing is low} problem. Using a state-of-the-art hydrodynamical simulation, we show that this discrepancy reflects shortcomings in standard galaxy-halo connection models rather than tensions within the $\Lambda$CDM paradigm itself. Specifically, this problem results from ignoring a variety of galaxy formation effects, including assembly bias, segregation of satellite galaxies relative to dark matter, and baryonic effects on the matter distribution. All these effects contribute towards overestimating gravitational lensing and, when combined, explain the amplitude and scale dependence of the {\it lensing is low} problem. We conclude that simplistic galaxy-halo connection models are inadequate to interpret clustering and lensing simultaneously, and that it is crucial to employ more sophisticated models for the upcoming generation of large-scale surveys.
\end{abstract}



\section{Introduction}
\label{sec:intro}

According to our best structure formation models, dark matter and baryons collapse together into halos within which baryons then cool, condense, and fragment to form galaxies \citep[e.g.,][]{white78}. We thus expect a well-defined ``galaxy-halo connection'' sensitive to fundamental aspects of the formation and evolution of structures. Two of the most precise avenues to study this connection are galaxy clustering (GC) and galaxy-galaxy lensing (GGL). The first refers to the spatial distribution of galaxies, and the second measures the deflection of light from background galaxies by matter surrounding foreground galaxies \citep[e.g.,][]{tyson1984_GalaxyMassdistribution, Miraldaescude91, brainerd1996_WeakGravitationalLensing, hudson1998_GalaxyGalaxyLensingHubble}. Notably, the combination of these observables enables precise measurements of the strength and scale-dependence of galaxy bias, cosmological parameters, and even the law of Gravity \citep[e.g.,][]{guzik2010_TestsGravityimaging, duncan2014ComplementarityGalaxyClustering, wibking2019_EmulatingGalaxyclustering, salcedo2022ExploitingNonlinearScales}.

Modern galaxy surveys sample large cosmic volumes with great precision, which enables detailed studies about the galaxy-halo connection. Strikingly, multiple analyses show that GGL measurements around massive galaxies are significantly smaller than predictions from theoretical models fitting their clustering; this is commonly known as the \lil problem. \citet{leauthaud2017_LensingLowcosmology} discovered this tension by analysing the clustering of galaxies from the CMASS sample of the Baryon Oscillation Spectroscopic Survey \citep[BOSS,][]{eisenstein11, dawson13} and lensing measurements around these galaxies from the Canada France Hawaii Telescope (CFHT) Lensing Survey \citep[CFHTLenS,][]{heymans2012_CFHTLenSCanadaFranceHawaiiTelescope, miller2013_BayesianGalaxyshape} and CFHT Stripe 82 survey \citep[CS82,][]{erben2013_CFHTLenSCanadaFranceHawaiiTelescope}. Using halo occupation distribution \citep[HOD, e.g.,][]{benson2000_NatureGalaxybias, peacock00, scoccimarro01} and subhalo abundance matching \citep[SHAM, e.g.,][]{vale04, conroy06, reddick13, contreras15, chavesmontero16, contreras2021FlexibleSubhaloAbundance} models, \citet{leauthaud2017_LensingLowcosmology} found that GGL measurements around CMASS galaxies were from 20 to 40\% lower than predictions from these models, and that this discrepancy was increasingly smaller for larger scales. Later, \citet[][\citetalias{lange2019_NewPerspectivesBOSS} hereafter]{lange2019_NewPerspectivesBOSS} corroborated these findings for other samples of the BOSS survey using a standard HOD model.

Perhaps surprisingly, the extent and scale-dependence of the \lil problem varied across subsequent studies, even for those analysing the same observational data \citep{wibking2020_CosmologyGalaxygalaxy, yuan2020_CanAssemblybias, yuan2021EvidenceGalaxyAssembly, lange2021HalomassRadialScalea, yuan2022ABACUSHODHighlyEfficient, yuan2022StringentS8Constraints}. However, the most statistically-significant result as of today confirms early findings. Using a standard HOD model, BOSS galaxies, and lensing data from the Dark Energy Survey year 3 data release \citep[DES-Y3;][]{amon2022_DarkEnergySurvey, secco2022_DarkEnergySurvey}, the fourth Kilo-Degree Survey data release \citep[KIDS-1000;][]{asgari2021_KiDS1000CosmologyCosmic}, and the Subaru Hyper Suprime-Cam survey year 1 data release \citep[HSC-Y1;][]{hikage2019_CosmologyCosmicsheara}, \citet[][\citetalias{amon2022ConsistentLensingClustering} hereafter]{amon2022ConsistentLensingClustering} found a $20-30\%$ small-scale discrepancy progressively decreasing towards larger scales.

There have been multiple attempts to understand the origin of this problem. A popular interpretation is that it is another face of the tension between growth of structure measurements from the early and late Universe \citep[e.g.,][]{amon2022_DarkEnergySurvey}, and therefore a consequence of a more profound inconsistency in the cosmological model. This explanation was motivated by the decrease in the tension when using theoretical models assuming $\sigma_8$ and $\Omega_\mathrm{m}$ values lower than those preferred by the analysis of {\it Planck} data \citep{leauthaud2017_LensingLowcosmology, lange2019_NewPerspectivesBOSS, wibking2020_CosmologyGalaxygalaxy, amon2022ConsistentLensingClustering, yuan2022StringentS8Constraints}. For example, \citetalias{amon2022ConsistentLensingClustering} showed that assuming cosmological parameters consistent with the analysis of lensing data from the KIDS and DES surveys significantly alleviates the tension on small scales and resolves it on large scales.

Another venue to decrease the \lil tension is to invoke galaxy formation physics. Traditional HOD models were designed to reproduce the clustering of luminosity-selected galaxies, and thus these might not be flexible enough to describe GC and GGL simultaneously, especially considering the complex galaxy selection criteria of most spectroscopic surveys. Along these lines, multiple studies showed that even when assuming the {\it Planck} cosmology, the \lil problem is significantly alleviated after accounting for the dependence of GC on halo properties other than halo mass \citep[i.e., galaxy assembly bias,][]{yuan2020_CanAssemblybias, yuan2021EvidenceGalaxyAssembly, yuan2022ABACUSHODHighlyEfficient, yuan2022StringentS8Constraints, amon2022ConsistentLensingClustering} or baryonic effects \citepalias{amon2022ConsistentLensingClustering} consistent with observational constraints from the kinematic Sunyaev-Zel'dovich effect \citep{amodeo2021AtacamaCosmologyTelescope}.

Nonetheless, standard galaxy-halo connection models typically neglect these effects, and their combined influence has yet to be assessed in state-of-the-art galaxy formation models. In this work, we study the origin of the \lil using the largest hydrodynamical simulation of the IllustrisTNG suite \citep{Pillepich2018a}, which reproduces an extensive range of observables such as the stellar mass function, galaxy colours, and the clustering of star-forming and quenched galaxies. First, we select a sample of galaxies from the IllustrisTNG simulation mimicking the properties of BOSS galaxies. Then, we compare GGL measurements around these galaxies with predictions from a standard HOD model optimised to reproduce their clustering, finding a discrepancy ranging from 25\% on small scales to 5\% on large scales, i.e., we reproduce the \lil problem in the IllustrisTNG simulation. We track the origin of this tension to the inadequacy of common assumptions of HOD models, finding that standard HOD implementations fail to capture multiple galaxy formation effects predicted by the IllustrisTNG simulation. Specifically, to accommodate the impact of these effects on GC, HOD models predict an incorrect galaxy occupation distribution, which causes the \lil problem. Notably, the extent and scale dependence of the resulting tension agrees remarkably well with that found in observational studies.

The structure of this paper is as follows. We start presenting the IllustrisTNG simulation and the numerical methods employed in this study in \S\ref{sec:methods}. In \S\ref{sec:btng}, we select IllustrisTNG galaxies mimicking the properties of BOSS galaxies and study the \lil problem for this sample. In \S\ref{sec:hod}, we quantify the validity of standard assumptions of HOD models for mock galaxies, and in \S\ref{sec:lil} we show how the inaccuracy of these assumptions generates the \lil problem. We summarise our findings and conclude in \S\ref{sec:conclusions}.


\section{Methods}
\label{sec:methods}

In this section, we present the dataset and numerical techniques we employ. In \S\ref{sec:methods_galaxies}, we describe the IllustrisTNG simulation, and in \S\ref{sec:methods_baryons} our approach for modelling the impact of baryonic effects on the matter distribution. Then, we detail how we compute GC and GGL from mock data in \S\ref{sec:methods_observables}. Finally, we describe our HOD implementation in \S\ref{sec:methods_hod}, and an emulator to accelerate the inference of the best-fitting HOD parameters to GC in \S\ref{sec:methods_emulator}.


\subsection{IllustrisTNG simulation}
\label{sec:methods_galaxies}

To carry out our calculations, we extract galaxy samples from cosmological simulations of the IllustrisTNG suite\footnote{\url{https://www.tng-project.org/}} \citep{Pillepich2018a, Pillepich2018b, Nelson2018a, Marinacci2018, Naiman2018, Springel2018, Nelson2018b}, which was carried out using the moving-mesh code \textsc{arepo} \citep{Springel2010}. This code solves for the joint evolution of dark matter, gas, stars, and supermassive black holes by incorporating a comprehensive galaxy formation model with star formation, radiative gas cooling, chemical enrichment, galactic winds, and stellar and AGN feedback \citep{Weinberger2017, Pillepich2018b}. 

We use publicly available data from the largest hydrodynamical simulation of the suite and its gravity-only version, TNG300-1 and TNG300-1-Dark, respectively. TNG300-1 evolved $2500^3$ gas tracers and the same number of dark matter particles in a periodic box of $205\,\Mpch$ on a side under Planck 2015 cosmology \citep{planck15XIII}. On the other hand, TNG300-1-Dark was run using the same initial conditions and configuration as TNG300-1 but {\it only} considering gravitational interactions. The mass resolution of dark matter and gas tracers is 4.0 and (initially) $0.7\times10^7\Msunh$ for TNG300-1, respectively, and $4.7\times10^7\Msunh$ for mass particles in TNG300-1-Dark. In what follows, we refer to these simulations as \tng and \dtng for simplicity. Unless otherwise stated, we estimate the matter density field for the \tng simulation using the stellar, baryonic, and dark matter components.

The IllustrisTNG simulation uses a standard friends-of-friends group finder with linking length $b=0.2$ to identify dark matter halos \citep{davis85}, and the \textsc{subfind} algorithm to identify self-bound structures within halos \citep{springel01, dolag09}, which are commonly known as subhalos. These algorithms assign to (sub)halos the coordinates of the (sub)halo particle with the minimum gravitational potential energy and peculiar velocities given by the sum of the mass-weighted velocities of all (sub)halo particles. It is standard to refer to subhalos located at the potential minimum of their host halos as centrals, other subhalos as satellites, and any subhalo with a stellar component as a galaxy. By construction, central and satellite galaxies present the same phase-space coordinates as their central and satellite subhalos.

Our analyses require identifying the counterparts of \tng structures in the \dtng simulation. There are two publicly-available catalogues for linking (sub)halos between these simulations: the first based on \textsc{lhalotree} \citep{nelson2015_IllustrisSimulationPublic} and the second on \textsc{sublink} \citep{Rodriguez_Gomez2015}. For each \tng subhalo, \textsc{sublink} finds the \dtng subhalo containing the largest fraction of its dark matter particles; this procedure can be done because \tng and \dtng employ the same initial conditions. \textsc{lhalotree} carries out the same procedure also beginning from \dtng subhalos, and it only confirms the links that agree when starting from both simulations. Therefore, the \textsc{sublink} catalogue presents larger completeness than the \textsc{lhalotree} catalogue, but the links of \textsc{sublink} are less robust than those from \textsc{lhalotree}.

We combine these two catalogues to increase completeness while keeping contamination as low as possible. To do so, we first store all subhalos whose link coincides for both algorithms. For those with a different match, we store the \textsc{sublink} matching when the relative difference between the value of $\vpeak$ of counterparts is smaller than 0.5 dex. Finally, if we find more than one \dtng subhalo satisfying this criterion for a particular \tng subhalo, we select the pair with the closest $\vpeak$ and $\Mpeak$. Following this procedure, we end up with 99.8 and 83.9\% of matches for central and satellite subhaloes hosting galaxies more massive than $\logMstar=10$, respectively, with an increasing success rate for more massive systems. This trend is explained by the finite mass resolution of the simulations, mass loss due to satellite-host interactions, slight differences in the timing at which mergers happen, and the difficulty in identifying and keeping track of satellites in high-density regions.


\subsection{Modelling baryonic effects}
\label{sec:methods_baryons}

Hydrodynamical simulations consistently predict that baryonic effects decrease the small-scale clustering of matter; however, the strength of this suppression varies from a few up to 40\% depending on the galaxy formation prescriptions implemented in each simulation \citep[e.g.,][]{Emberson2018, arico2020ModellingLargescaleMassa}. Observational constraints on the strength and scale dependence of baryonic effects are also uncertain \citep[e.g.,][]{amodeo2021AtacamaCosmologyTelescope, Chen:2022}; as a result, instead of simply considering the type of baryonic effects predicted by the \tng simulation, we model a plausible range of baryonic effects by evaluating the ``baryonification'' algorithm on top of the \dtng simulation \citep{schneider2015NewMethodQuantifya, schneider2019QuantifyingBaryonEffectsa, arico2020ModellingLargescaleMassa, arico2021SimultaneousModellingMatter}. 

This algorithm alters the position of mass tracers in gravity-only simulations to mimic the effect of star formation, feedback, and gas cooling on the mass distribution. Notably, using only a few free parameters, this technique successfully captures the impact of baryonic effects on the 2- and 3-point statistics of almost all publicly available hydrodynamical simulations \citep{schneider2020BaryonicEffectsWeak, arico2021SimultaneousModellingMatter}. In our analysis, we generate two ``baryonified'' \dtng simulations using baryonification parameters measured from the low- and high-AGN simulations of the BAryons and HAloes of MAssive Systems suite \citep[BAHAMAS,][]{mccarthy2017BAHAMASProjectCalibrated, mccarthy2018BAHAMASProjectCMBlargescale}, which use AGN feedback prescriptions weaker and stronger than the standard BAHAMAS run, respectively. We consider these simulations because they bracket the range of baryonic effects predicted by the majority of state-of-the-art simulations \citep[e.g.,][]{arico2020ModellingLargescaleMassa}.


\subsection{Galaxy clustering \& galaxy-galaxy lensing}
\label{sec:methods_observables}

We characterise GC using the projected correlation function, $\omega_{\rm p}(r_\perp)$, which provides the excess probability of finding a galaxy at perpendicular to the line-of-sight distance $r_\perp$ relative to expectations for a random sample. We compute this observable by integrating the three-dimensional two-point correlation function, $\xi_\mathrm{gg}$, along the line of sight
\begin{equation}
    \omega_{\rm p}(r_\perp) = 
    \int_{-s_\parallel^\mathrm{max}}^{s_\parallel^\mathrm{max}} \xi_\mathrm{gg}(r_\perp, s_\parallel)\,\mathrm{d}s_\parallel,
\end{equation}
where $s_\parallel = r_\parallel + (1+z_\mathrm{box})\,v_\parallel/H(z_\mathrm{box})$ is the line-of-sight redshift-space coordinate, $v_\parallel$ is the peculiar radial velocity, $H$ is the Hubble parameter, $z_\mathrm{box}$ is the redshift of the simulation box, and $s_\parallel^\mathrm{max}$ is the integration boundary.

A foreground mass distribution induces a shear signal on background sources that depends upon the transverse and parallel distances between the lens-source pair. This distortion is proportional to the excess surface density
\begin{equation}
    \Delta\Sigma(r_\perp) = \overline{\Sigma}(\leq r_\perp) - \Sigma(r_\perp),
\end{equation}
where $\Sigma$ is the azimuthally-averaged surface mass density, and $\overline{\Sigma}(\leq r_\perp)$ is the mean surface density within a projected radius $r_\perp$. We estimate the mean surface mass density using
\begin{equation}
    \Sigma(r_\perp) = \Omega_\mathrm{m}\,\rho_\mathrm{crit} 
    \int_{-r_\parallel^\mathrm{max}}^{r_\parallel^\mathrm{max}} \xi_\mathrm{gm}(r_\perp, r_\parallel)\,\mathrm{d}r_\parallel,
\end{equation}
where $r_\parallel$ refers to the projected distance along the line of sight, $r_\parallel^\mathrm{max}$ is the integration boundary, $\xi_\mathrm{gm}$ is the galaxy-matter three-dimensional cross-correlation function, $\Omega_\mathrm{m}$ and $\rho_\mathrm{crit}$ are the matter and critical density of the Universe, respectively, and the azimuthally-averaged surface mass density within a radius $r_\perp$ is
\begin{equation}
    \overline{\Sigma}(\leq r_\perp) = \frac{2}{r_\perp^2}\int_0^{r_\perp} \Sigma(\tilde{r})\,\tilde{r} \,\mathrm{d}\tilde{r}.
\end{equation}

Operationally, we compute $\omega_{\rm p}$ and $\Delta\Sigma$ by first measuring $\xi_{\rm gg}$ and $\xi_{\rm gm}$ using a combination of routines from the high-performance python package \textsc{corrfunc} \citep{sinha2020_CORRFUNCSuiteblazing} and our own. Specifically, we use 13 logarithmically-spaced bins between $r_\perp=0.1$ and $25\,\Mpch$, integration boundaries $s_\parallel^\mathrm{max}=r_\parallel^\mathrm{max}=30\,\Mpch$, and we compute the average of each observable after considering the three simulation coordinate axes as line of sights. We do not consider larger scales or integration boundaries because the results become too noisy due to the limited size of the \tng simulation box. 

On scales larger than $r=0.1\,\Mpch$, the impact of baryonic physics on the matter correlation function is the same in the TNG300-1 and TNG100-1 simulations, where the second is a simulation of the IllustrisTNG suite with $\simeq8$ times higher resolution than the TNG300-1 simulation \citep{Springel2018}. We thus conclude that the resolution of \tng is enough for measuring GGL across the entire range of scales considered. In practice, we measure \ds using a subsampled version of the \tng and \dtng matter density fields diluted by a factor of 1000; we checked that this dilution enables sub-percent measurements down to $r_\perp=0.1\,\Mpch$.


\subsection{HOD model}
\label{sec:methods_hod}

Halo Occupation Distribution models \citep[e.g.,][]{berlind2002_HaloOccupationDistribution, kravtsov04, zheng05} are among the most widely used methods for modelling GC and GGL. In particular, \lil studies use HOD formulations ranging from the simplest ones \citep[][]{leauthaud2017_LensingLowcosmology, lange2019_NewPerspectivesBOSS, amon2022ConsistentLensingClustering} to extensions modelling physical effects such as satellite segregation and assembly bias \citep[e.g.,][]{yuan2022StringentS8Constraints}. Throughout this work, we study the \lil problem using a HOD formation very similar to those used in \citetalias{lange2019_NewPerspectivesBOSS} and \citetalias{amon2022ConsistentLensingClustering}.

The primary quantity of interest in HOD models is $P(N_\mathrm{gal}|M_\mathrm{h})$, which provides the probability of a halo of mass $M_\mathrm{h}$ to host $N_\mathrm{gal}$ galaxies. Following \citet{zheng05}, we model the occupation number of central and satellite galaxies separately. We consider that the average occupation of central galaxies is described by
\begin{equation}
    \label{eq:hod_cen}
    \langle N_\mathrm{cen}|M_\mathrm{h} \rangle = \frac{1}{2} + \frac{1}{2} \mathrm{erf}\left(\frac{\log M_\mathrm{h} - \log M_\mathrm{min}}{\sigma_{\log M}}\right),
\end{equation}
where $M_\mathrm{min}$ refers to the characteristic minimum mass of a halo hosting a central galaxy, $\sigma_{\log M}$ indicates the width of the transition from zero probability to unity, and $\mathrm{erf}$ denotes the error function,
\begin{equation}
    \mathrm{erf}(y) = \frac{2}{\sqrt{\upi}}\int_0^y e^{-x^2} \mathrm{d}x.
\end{equation}

For satellites, we use
\begin{equation}
    \label{eq:hod_sat}
    \langle N_\mathrm{sat}|M_\mathrm{h} \rangle =
    \left\{\begin{aligned}
        0\quad &\,\mathrm{if}\,M_\mathrm{h}\leq M_0,\\
        \left(\frac{M_\mathrm{h}-M_0}{M_1}\right)^\alpha &\,\mathrm{if}\,M_\mathrm{h}>M_0,
       \end{aligned}
 \right.
\end{equation}
where $M_0$ is the mass threshold below which a halo does not host any satellite, $M_1$ is the mass for which a halo is expected to host approximately one satellite galaxy, and $\alpha$ controls the steepness of the increase in the number of satellite galaxies with halo mass. Contrary to some HOD implementations, we decouple the central and satellite probabilities because selection criteria may result in halos hosting only satellite galaxies (see \S\ref{sec:hod_parametric}).

By definition, a halo only contains at most one central galaxy, which suggests using a nearest integer distribution with mean $\langle N_\mathrm{cen}|M_\mathrm{h} \rangle$ for populating halos with central galaxies. The situation is more complicated for satellite galaxies as their number is not bound between 0 and 1; the standard approach is to consider a Poisson distribution with mean $\langle N_\mathrm{sat}|M_\mathrm{h} \rangle$ \citep[see also, e.g.,][]{jimenez2019_ExtensionsHalooccupation, avila2020CompletedSDSSIVExtended, hadzhiyska2022MillenniumTNGProjectRefiningb}. Once we populate halos using these probability distributions, we assign to central and satellite galaxies the phase-space coordinates of their host halos and randomly selected dark matter particles from their host halos, respectively.


\subsection{HOD emulator}
\label{sec:methods_emulator}

In subsequent sections, we optimise the free parameters of our HOD implementation to jointly reproduce GC and number density measurements from mock galaxies. We describe our approach to do so next.

To speed up the inference of HOD parameters, we create surrogate models predicting GC and number density as a function of the value of HOD parameters. First, we design a hypercube in the 5-dimensional parameter space of our HOD implementation,
\begin{equation}
\left\{
\begin{aligned}
    &\log_{10}(M_{\rm min}[\Msunh])   \in [12, 14],\\
    &\sigma_{\log M}                  \in [0.01, 1.75],\\
    &\log_{10}(M_0[\Msunh])           \in [12, 15],\\
    &\log_{10}(M_1[\Msunh])           \in [12.5, 15],\\
    &\alpha                           \in [0.5, 2.5],\\
\end{aligned}   
\right.
\end{equation}
and then we sample it according to a Latin-Hypercube using 250\,000 points. For each of these points, we populate the \dtng simulation using our HOD model, and then we measure the number density and projected clustering of the resulting sample (see \S\ref{sec:methods_observables}). Using different random draws, we repeat this procedure 100 times for each point to reduce stochastic noise owing to the probabilistic nature of HOD models. Note that we populate the \dtng simulation to ensure that the impact of cosmic variance on \tng and HOD predictions is analogous.

To predict these observables, we built a pair of fully-connected neural networks using \textsc{pytorch} \citep{pytorch}. Each neural network presents input and output layers with as many neurons as data values (1 and 13 for number density and projected clustering, respectively), 4 hidden layers with 4 neurons per input data value, and SELU activation functions \citep{klambauer2017SelfNormalizingNeuralNetworks} for all layer but the last one, for which we use a linear activation. To train these networks, we employ 90\% of the previous measurements, ``Xavier'' weight initialisation \citep{glorot2010UnderstandingDifficultyTraining}, the AMSGrad variant of the Adam optimisation algorithm \citep{kingma2014AdamMethodStochastic, reddi2022ConvergenceAdam}, and a mean absolute error loss function. Using the remaining 10\% of the previous measurements, we checked that the accuracy of both networks is approximately $1\%$ over all scales considered. Each emulator roughly takes $1\,\mathrm{ms}$ per evaluation and $200\,\mathrm{ms}$ for batches of 100\,000 evaluations.

We optimise HOD parameters to describe the clustering of a target galaxy sample using the publicly available Affine Invariant Markov chain Monte Carlo (MCMC) Ensemble sampler {\sc emcee} \citep{foremanmackey13}\footnote{\url{https://emcee.readthedocs.io/en/stable/}}. For each step of the Markov chain, {\sc emcee} first draws a value for each HOD parameter. Then, we evaluate the previous two surrogate models for this combination of parameters to obtain the number density and GC that the HOD model would predict. To estimate the probability of these parameters, we compare the predicted number density and GC with those measured from the target galaxy sample. To do so, we use a Gaussian likelihood and a diagonal covariance matrix with the number density element set to 0.05 dex and the GC elements set to a scaled version of the BOSS covariance matrix. Specifically, we scale these elements so the average signal-to-noise of GC for the target sample is 40, which ensures a rapid convergence of MCMC chains. This simple approach suffices since we are only interested in the best-fitting HOD solution and not in the precise shape of the posterior or correlations among parameters. We checked that best-fitting solutions are largely insensitive to slight variations in the amplitude and scale dependence of clustering terms.

We run {\sc emcee} for each target sample using 300 independent chains of 800 steps, a burn-in phase of 150 steps, and uniform priors matching the hypercube used to train the emulators. We verify that this configuration results in a robust sampling of the parameter posterior.


\section{\bosstng catalogue}
\label{sec:btng}

In this section, we first generate a \tng mock galaxy catalogue that mimics the properties of BOSS galaxies. Then, for this sample, we test the consistency of GGL measurements and predictions from a HOD model optimised to reproduce its clustering.


\subsection{Sample selection and properties}
\label{sec:btng_selection}

\begin{figure}
    \centering
    \includegraphics[width=\columnwidth]{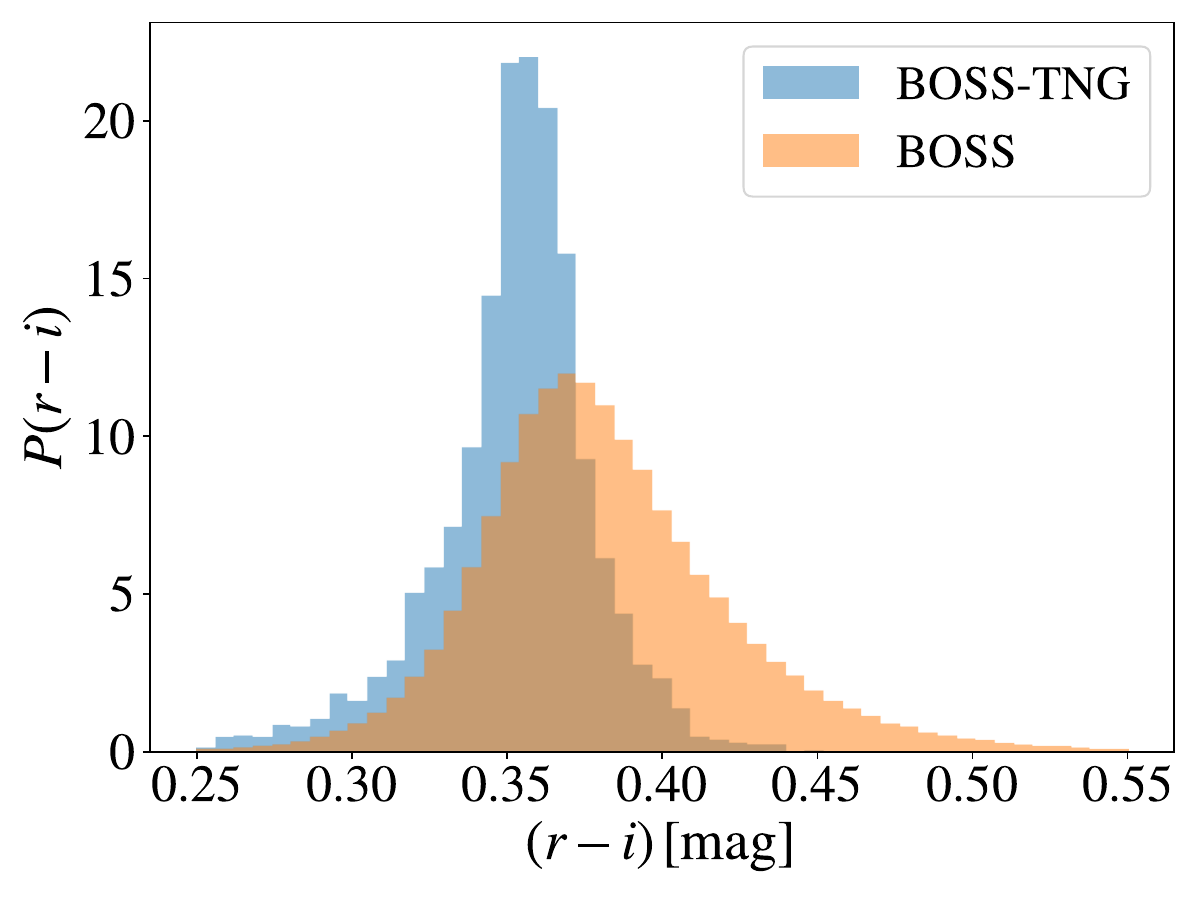}
    \includegraphics[width=\columnwidth]{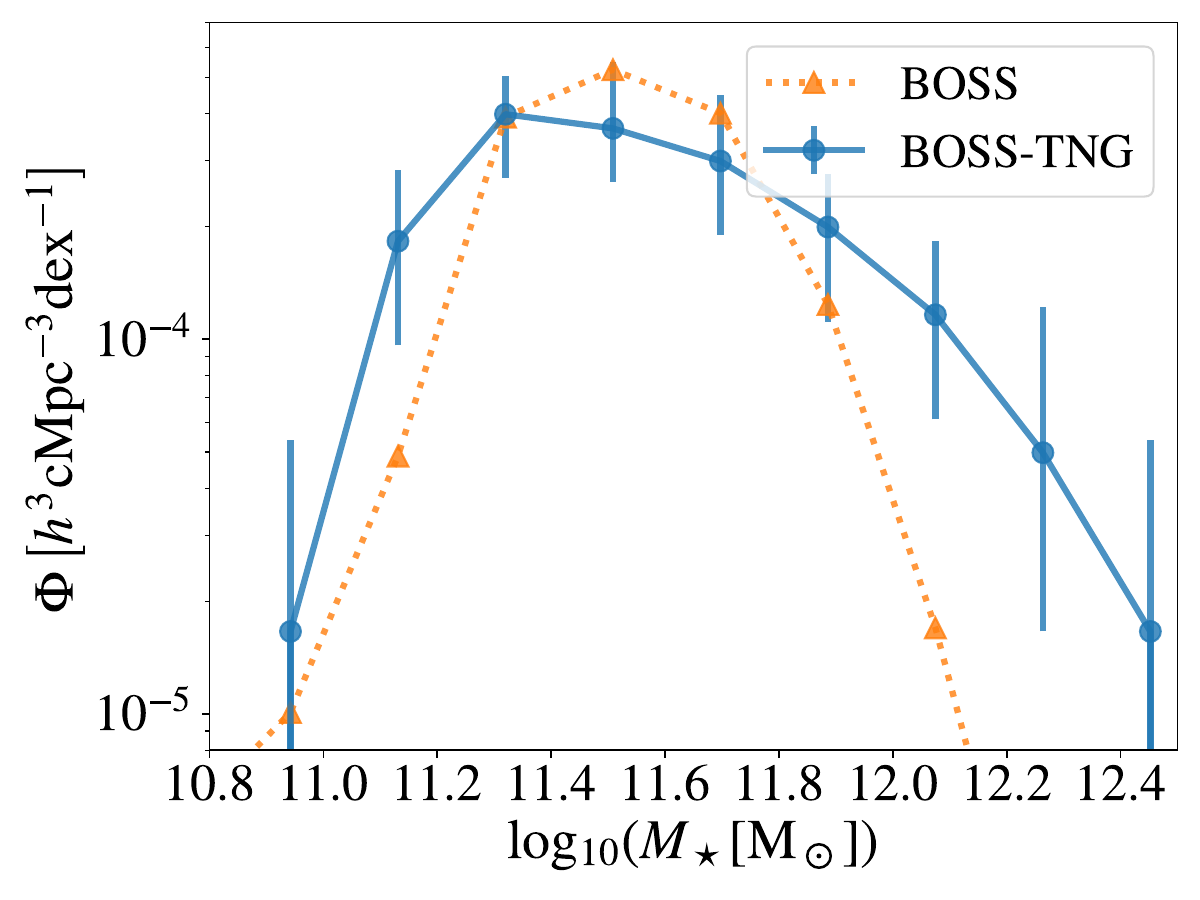}
    \caption{
    Rest-frame $r-i$ colours (top panel) and galaxy stellar mass function (bottom panel) of \bosstng and BOSS galaxies. The blue colour indicates the results for \bosstng galaxies, whereas the orange colour does so for BOSS galaxies. Error bars denote uncertainties due to the combination of cosmic variance, sample discreteness, observational errors, and model shortcomings. We find a broad agreement between the properties of \bosstng and BOSS galaxies.
    }
    \label{fig:boss_props}
\end{figure}

The analysis of GC and GGL measurements from BOSS galaxies provides the best observational estimates of the \lil problem. Out of these GGL measurements, those around galaxies from the low redshift sample of the BOSS survey \citep[LOWZ;][]{eisenstein2001_SpectroscopicTargetSelection}, which targeted luminous red galaxies (LRGs) at $z<0.4$, show the greatest consistency among lensing surveys \citep{leauthaud2022LensingBordersBlind}. Motivated by this, we investigate the origin of the \lil problem using a galaxy sample from the \tng simulation mimicking the properties of LOWZ galaxies. Note that the \tng simulation captures the colours \citep{Nelson2018a} and clustering \citep{Springel2018} of LRGs with remarkable precision and presents enough volume for statistically-significant studies.

To build this mock catalogue, the \bosstng sample in what follows, we first transform the publicly-available, rest-frame magnitudes of \tng galaxies at $z=0$ to observed-frame fluxes at $z=0.3$, which approximately corresponds to the median redshift of LOWZ galaxies. We use $z=0$ data because the number of LOWZ-like galaxies inside the \tng box at $z=0.3$ is quite limited, making GC and GGL measurements too noisy. Nevertheless, this election should not affect our findings as observational studies find that the \lil problem is approximately redshift independent \citepalias[e.g.,][]{lange2019_NewPerspectivesBOSS, amon2022ConsistentLensingClustering}. We further discuss this issue in \S\ref{sec:lil_origin}.

We continue by perturbing observed-frame fluxes with Gaussian errors reproducing the level of photometric uncertainties affecting LOWZ galaxies. After that, we measure the position of \tng galaxies along and across the red sequence in the \{$g-r$, $r-i$\} plane, which we identify by performing a linear fit to the colours of quenched galaxies with $\logMstar>9.5$:
\begin{equation}
\left\{
\begin{aligned}
c_\perp &= (r-i) - 0.38(g-r) - 0.074, \\
c_\parallel &= (r-i) + 2.65(g-r) + 1.70,
\end{aligned}\label{eq:colors_rot}
\right.
\end{equation}
where $c_\perp$ and $c_\parallel$ are the parallel and perpendicular distances to the red sequence, respectively. Finally, we select \tng galaxies satisfying LOWZ-like, colour-based selection criteria \citep{eisenstein2001_SpectroscopicTargetSelection},
\begin{equation}
\left\{
\begin{aligned}
|c_\perp| &< 0.008, \\
r &< 3 c_\parallel + r_n,\\
r &< 20,\\
\sigma_g &< 0.2,\\
\sigma_r &< 0.1,\\
\sigma_i &< 0.1,
\end{aligned}\label{eq:colors_sel}
\right.
\end{equation}
where $\sigma_g$, $\sigma_r$, $\sigma_i$ are the magnitude uncertainty in the $g$, $r$, and $i$ bands, respectively. The first criterion ensures that we only select galaxies that belong to the red sequence, the second sets the luminosity threshold of the sample via the free parameter $r_n$, and the remainder guarantee the selection of galaxies with sufficient signal-to-noise ratio. 

We optimise the value of $r_n$ to select \bosstng galaxies with similar number density as LOWZ galaxies, finding that $r_n=8.5$ mag results in a number density of $\log_{10} n=-3.5 \, h^{3}{\rm Mpc^{-3}}$, which is approximately the same as that of LOWZ galaxies \citep{parejko2013_ClusteringGalaxiesSDSSIII}. We check that \bosstng galaxies and their host halos are well-resolved in the \tng simulation: the first and second present more than 3\,200 stellar and 18\,500 dark matter particles in all cases, respectively. Finally, we find that the satellite fraction of \bosstng galaxies is similar to that of LOWZ galaxies: 18.6 and $12\pm2\%$ for the first and second \citep{parejko2013_ClusteringGalaxiesSDSSIII}, respectively. 

In the top panel of Fig.~\ref{fig:boss_props}, we display the rest-frame $r-i$ colour distribution of \bosstng and BOSS galaxies. The blue colour indicates the results for \bosstng, whereas the orange colour does so for LOWZ galaxies in the Northern Galactic Cup (NGC). We apply $k$ and evolutionary corrections to BOSS rest-frame colours that we compute following a similar approach as \citet{blanton2007KCorrectionsFilterTransformations}. Overall, we find good agreement between observed and simulated galaxies. Nevertheless, BOSS galaxies present on average redder colours than \bosstng galaxies, and these colours span a more extensive range for the first sample. The origin of this discrepancy is that the location of the \tng red sequence is slightly shifted towards bluer colours relative to observations and its width is too narrow at the high-mass end \citep{Nelson2018a}.

In the bottom panel of Fig.~\ref{fig:boss_props}, we display the stellar mass function of \bosstng and BOSS galaxies. For \bosstng galaxies, we use stellar masses measured inside a radius within which the surface brightness profile is more luminous than 20.7 mag $\mathrm{arcsec}^{-2}$ in the $K$ band. For NGC-LOWZ galaxies, we use Granada stellar masses \citep{Ahn2014} measured assuming a wide prior on the star-formation history, possibility for dust extinction, and a \citet{Kroupa2001} initial mass function\footnote{See \citet{maraston2013StellarMassesSDSSIII, bundy2017Stripe82Massive} for more information about BOSS stellar masses.}. Error bars show the result of adding in quadrature the most significant sources of uncertainty affecting the stellar mass function: cosmic variance, sample discreteness, observational errors, and model shortcomings. We estimate the impact of cosmic variance by first splitting the simulation box into 8 equally-sized subvolumes, and then computing the dispersion between measurements in each subvolume. We consider Poisson errors to account for sample discreteness, and we estimate the influence of observational errors and model shortcomings on stellar mass estimates by perturbing the logarithm of \bosstng stellar masses according to 0.2 dex Gaussian errors \citep[e.g.,][]{Lower2020}.

We find a good agreement between the stellar mass function of \bosstng and BOSS galaxies at the low-mass end: both peak at approximately the same value and plummet for lower masses than the peak; the LOWZ selection criteria originate this characteristic trend \citep[e.g.,][]{leauthaud2016_Stripe82Massive, tinker2017_CorrelationHaloMass}. On the other hand, the stellar mass functions depart at the high-mass end. We find that this discrepancy depends strongly upon the approach to measure stellar masses in the \tng simulation, and it is likely caused by a combination of cosmic variance and an inconsistent comparison between stellar masses from observations and the simulation.


\subsection{The \textbf{\textit{lensing is low}} problem}
\label{sec:btng_lil}

\begin{figure}
    \centering
    \includegraphics[width=\columnwidth]{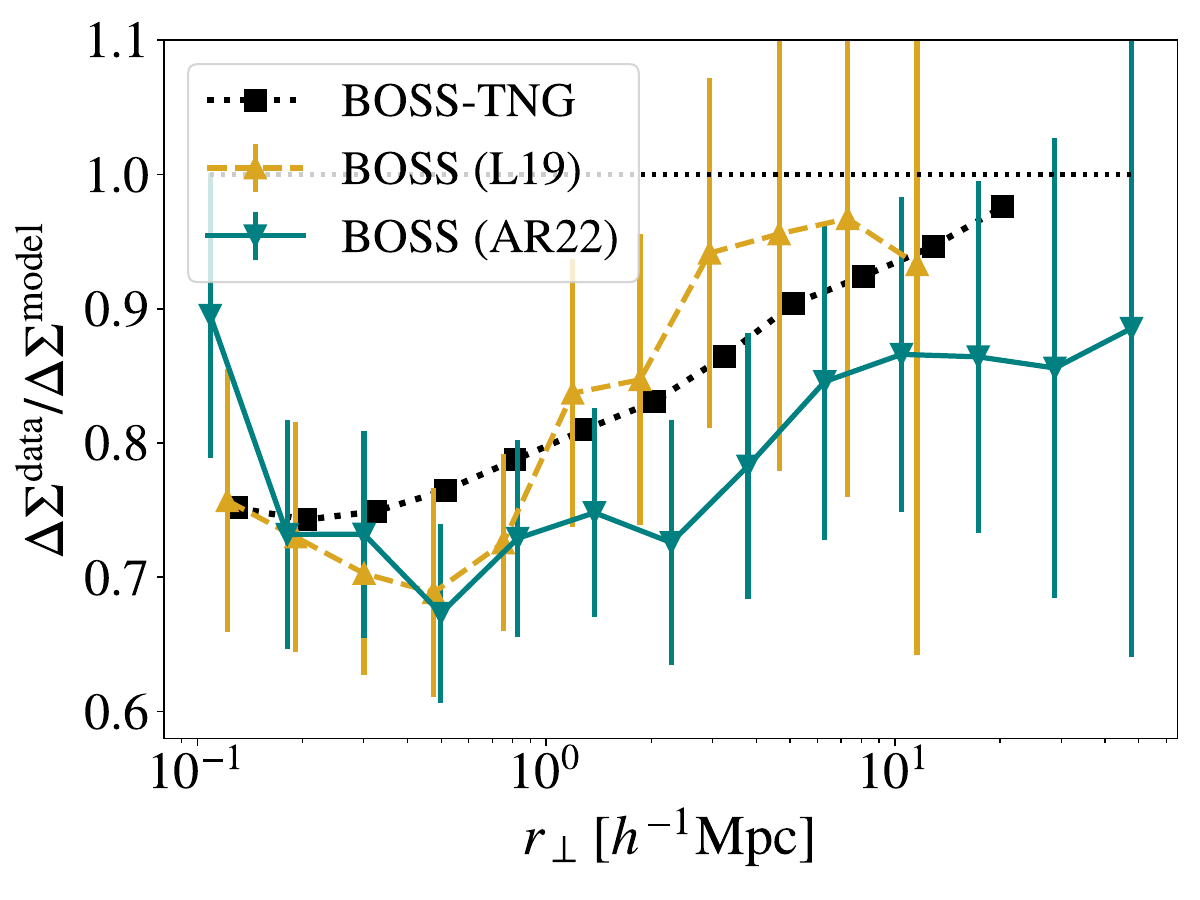}
    \caption{
    Ratio between lensing measurements around massive galaxies and predictions from best-fitting HOD models to their clustering. The departure of this ratio from unity is commonly known as the \lil problem. The dotted line indicates the results for \bosstng galaxies, the yellow line for BOSS galaxies and CFHTLenS lensing data \citepalias{lange2019_NewPerspectivesBOSS}, and the green line for BOSS galaxies and the combination of KiDS and DES lensing data \citepalias{amon2022ConsistentLensingClustering}. Error bars denote 1-$\sigma$ uncertainties. As we can see, the \tng simulation predicts a \lil problem with similar strength and scale dependence as observational estimates.
    }
    \label{fig:boss_lil}
\end{figure}

It is standard to quantify the \lil problem by the ratio between measurements of the excess surface density around a set of galaxies and theoretical expectations for this quantity from best-fitting models to their clustering. Measurements and predictions should match because GC and GGL are sensitive to the same galaxy-halo connection, and thus the departure of this ratio from unity signals a tension between theory and observations. Therefore, it is essential to note that the ratio between $\ds$ measurements and theoretical expectations is independent of the strength of both GC and GGL. Consequently, we can study the \lil problem for different galaxy samples by comparing measurements of the aforementioned ratio.

To compute this ratio for the \bosstng sample, we first measure the excess surface density around \bosstng galaxies following the procedure explained in \S\ref{sec:methods_observables}. Then, we compute the value of the free parameters of our HOD implementation that provide the best-fitting solution to the projected clustering of \bosstng galaxies (see \S\ref{sec:methods_emulator}). After that, we evaluate our HOD implementation in \dtng simulation using these best-fitting parameters, populating it with galaxies. Finally, we compute the excess surface density around mock galaxies. Consequently, this measurement provides the GGL prediction of a HOD model optimised to reproduce the clustering of \bosstng galaxies. We iterate 40 times using different random seeds to decrease stochastic uncertainties associated with the probabilistic nature of HOD models.

In Fig.~\ref{fig:boss_lil}, the black dotted line shows the ratio between GGL measurements and average HOD predictions for \bosstng galaxies, whereas solid lines do so for BOSS galaxies when assuming the {\it Planck} cosmology. The yellow line displays average results from BOSS galaxies at six redshift bins between $z=0.1$ and 0.7 using CFHTLenS lensing data \citepalias{lange2019_NewPerspectivesBOSS}, whereas the green line shows average results from BOSS galaxies at three\footnote{We compute the average ratio without using data in the $z=0.43-0.54$ bin because \citetalias{amon2022ConsistentLensingClustering} claims that the results from this bin are contaminated by systematics.} redshift bins between $z=0.15$ and 0.7 using the combination of KiDS and DES lensing data \citepalias{amon2022ConsistentLensingClustering}. The only significant difference between our HOD implementation and those used by these observational studies is that we assign the position of randomly selected dark matter halo particles to satellite galaxies, whereas \citetalias{lange2019_NewPerspectivesBOSS} and \citetalias{amon2022ConsistentLensingClustering} assume that satellite galaxies follow a Navarro-Frenk-White profile \citep[NFW,][]{navarro96a}.

As we can see, the \tng simulation predicts \lil under the \lcdm cosmological model, suggesting that this tension does not arise from assuming incorrect cosmological parameters \citep[as suggested by some works, e.g.,][]{leauthaud2017_LensingLowcosmology, lange2019_NewPerspectivesBOSS, wibking2020_CosmologyGalaxygalaxy, amon2022ConsistentLensingClustering, yuan2022StringentS8Constraints} or it caused by extensions of the standard model \citep[e.g.,][]{leauthaud2017_LensingLowcosmology}. Notably, the strength and scale dependence of this effect for \bosstng and BOSS galaxies agree within uncertainties. \citetalias{lange2019_NewPerspectivesBOSS} and \citetalias{amon2022ConsistentLensingClustering} assume the same cosmology and analogous theoretical modelling, and thus differences between these studies are likely due to discrepancies between lensing estimates from distinct lensing surveys \citep[see][for a recent comparison]{leauthaud2022LensingBordersBlind}.

To further quantify this tension, we compute the weighted average of the previous ratios over all scales,
\begin{equation}
    \label{eq:lil_a}
    A \equiv 
    \left(\sum_i \sigma^{-2}_i\right)^{-1}
    \sum_i \sigma^{-2}_i \frac{\Delta\Sigma^\mathrm{data}(r_{\perp,i})}{\Delta\Sigma^\mathrm{model}(r_{\perp,i})},
\end{equation}
where $i$ goes through all radial bins used in the analysis and $\sigma_i$ is the uncertainty in the ratio $\Delta\Sigma^\mathrm{data}/\Delta\Sigma^\mathrm{model}$ at a scale $r_{\perp,i}$. We find $A=0.79\pm0.03$, $0.77\pm0.10$, and $0.76\pm0.03$ for \bosstng, \citetalias{lange2019_NewPerspectivesBOSS}, and \citetalias{amon2022ConsistentLensingClustering}, respectively, which are mutually consistent within statistical uncertainties. Due to the outstanding agreement between estimates from the \tng simulation and observations, we investigate the origin of this problem using \bosstng galaxies throughout the remainder of this work.


\section{Testing HOD assumptions}
\label{sec:hod}

In \S\ref{sec:btng}, we showed that the \tng simulation predicts a \lil problem with a magnitude and scale dependence similar to that found in observational studies. In this section, we qualitatively study whether the origin of this tension is related to the ability of galaxy-halo connection models to capture a variety of galaxy formation effects. Specifically, we quantify the precision of standard HOD models reproducing the following features of the galaxy-halo connection for \bosstng galaxies:

\begin{itemize}
    \item mass-dependent occupation distribution (\S\ref{sec:hod_parametric}),
    \item radial and anisotropic satellite distribution (\S\ref{sec:hod_sats} and \ref{sec:hod_slens}),
    \item assembly bias (\S\ref{sec:hod_ab}),
    \item concentration-dependent occupation distribution (\S\ref{sec:hod_conc}),
    \item redistribution of mass in halos due to baryonic effects (\S\ref{sec:hod_baryons}).
\end{itemize}

We note that some of these effects have been studied in hydrodynamical simulations using other galaxy samples \citep{beltz-mohrmann2020TestingAccuracyHalo, yuan2022IllustratingGalaxyhaloConnectiona}.


\subsection{Parametric occupation}
\label{sec:hod_parametric}

\begin{figure}
    \centering    
    \includegraphics[width=\columnwidth]{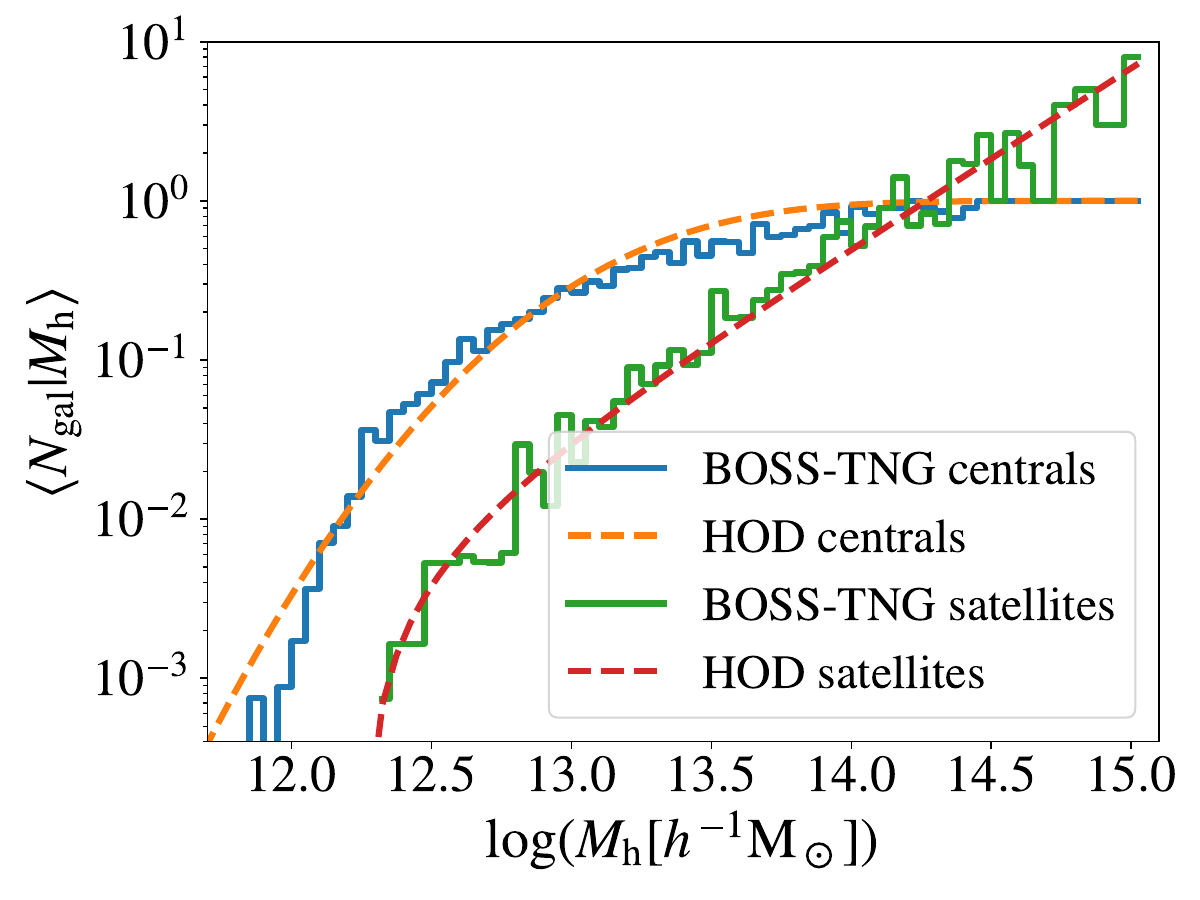}
    \caption{
    Mass-dependent halo occupation distribution of \bosstng galaxies. The blue and green solid lines indicate the results for central and satellite \bosstng galaxies, respectively, whereas the orange and red dashed lines do so for the best-fitting solution of standard parametric forms to these. Standard forms describe the satellite occupation of \bosstng galaxies accurately; however, these struggle to reproduce their central occupation.
    }
    \label{fig:hod_parametric}
\end{figure}

HOD models use simple parametric forms to describe the mass-dependent halo occupation distribution of central and satellite galaxies. Nonetheless, these forms were designed for stellar mass- and luminosity-selected galaxies \citep[e.g.,][]{zheng05}, and thus might not be precise enough for BOSS galaxies, which present colour-, magnitude-, and stellar mass-dependent incompleteness \citep[e.g.,][]{more2015_WeakLensingSignal, leauthaud2016_Stripe82Massive, rodriguez-torresClusteringGalaxiesSDSSIII2016, saito16}. In this section, we quantify the precision of standard parametric forms capturing the occupation distribution of \bosstng galaxies.

In Fig.~\ref{fig:hod_parametric}, blue and green solid lines display the average occupation distribution of central and satellite \bosstng galaxies, respectively, whereas orange and red dashed lines show the best-fitting solution of Eqs.~\ref{eq:hod_cen} and \ref{eq:hod_sat} to these distributions. We compute the best-fitting value of the parameters controlling these forms using the routine {\sc minimize} from {\sc scipy} \citep{virtanen2020_SciPyFundamentalalgorithms}. As we can see, the \bosstng satellite occupation increases with halo mass as a power law, which is precisely captured by the parametric model. The \bosstng central occupation also increases with mass, but it only reaches unity for very large halo masses due to the colour-dependent incompleteness of the sample. The central form struggles to capture this trend: it over- and under-predicts the fraction of galaxies in high and low-mass halos, respectively. Given that GGL increases with the average host halo mass of the galaxy sample, we expect standard HOD models to overestimate the magnitude of GGL for \bosstng galaxies.

It is worth noting that our HOD implementation decouples the probability of finding central and satellite galaxies in halos. However, some HOD models used in \lil studies assume that satellite galaxies can only reside in halos hosting a central galaxy \citepalias[e.g.,][]{amon2022ConsistentLensingClustering}. Interestingly, we find that \bosstng galaxies do not satisfy this assumption: 36\% of \bosstng satellites reside in halos not hosting a central galaxy. This fraction decreases with host halo mass and becomes negligible for halos more massive than $\logMh=14$. Therefore, HOD models considering this assumption could both overestimate the fraction of centrals in low-mass halos and underestimate the fraction of satellites in these. Since centrals reside in denser regions than satellites, this type of HOD implementation might overestimate GGL for \bosstng galaxies (see Appendix~\ref{app:small}).


\subsection{Satellite segregation}
\label{sec:hod_sats}

\begin{figure}
    \centering    
    \includegraphics[width=\columnwidth]{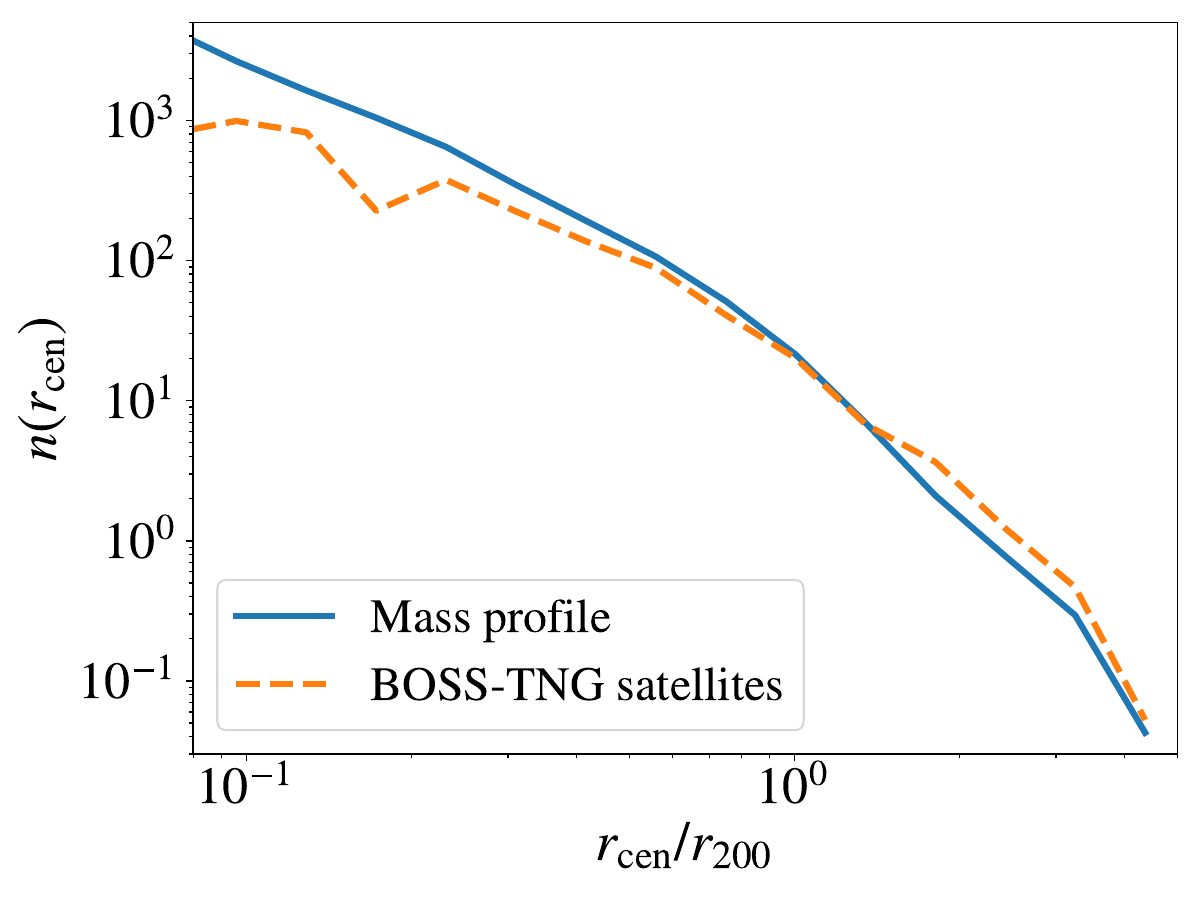}
    \caption{Mass and satellite profile of \bosstng halos. The blue solid line denotes the average radial distribution of dark matter particles in \tng halos weighted by the number of \bosstng satellites in each halo; the orange dashed line indicates the average radial profile of \bosstng satellites. As we can see, \bosstng satellites do not follow the mass profile in an unbiased fashion.
    }
    \label{fig:hod_radial}
\end{figure}

Standard HOD models assign to satellite galaxies coordinates based on the distribution of mass within their host halos. Nevertheless, satellites selected according to their star formation or colours may be segregated from the underlying mass distribution. For instance, star-forming galaxies preferentially sit on the outskirts of haloes because galaxies in the inner regions have already stopped forming stars due to ram pressure stripping, tidal stripping, or other effects \citep[e.g.,][]{Orsi:2018}. In this section, we investigate the accuracy of this assumption for \bosstng galaxies.

In Fig.~\ref{fig:hod_radial}, we display the number density of \bosstng satellites as a function of distance to the centre of their host halos, $n(r')=(3/4\pi r'^3) (\mathrm{d}N/{\rm d}r')$, where $N$ is the number of satellite galaxies, $r'=r_\mathrm{cen}/r_{200}$, $r_\mathrm{cen}$ is the distance of a satellite to the halo centre, and $r_{200}$ refers to the radius at which the halo density reaches 200 times the critical density of the Universe. The blue solid line indicates the average radial distribution of mass within \tng halos, which we weight by the number of \bosstng satellites in each halo so we can compare it to the \bosstng satellite profile. As we can see, the satellite profile is flatter than the mass profile.

To quantify this difference, we compute the best-fitting NFW solution to the satellite and mass profiles using {\sc minimize}. We find that the best-fitting satellite profile presents half the concentration of the best-fitting mass profile. As a result, we expect standard HOD implementations to place satellite galaxies closer to the halo centre than the average distance of \bosstng galaxies, thereby overestimating their small-scale GC and GGL.


\subsection{Subhalo lensing and halo triaxiality}
\label{sec:hod_slens}

\begin{figure}
    \centering    
    \includegraphics[width=\columnwidth]{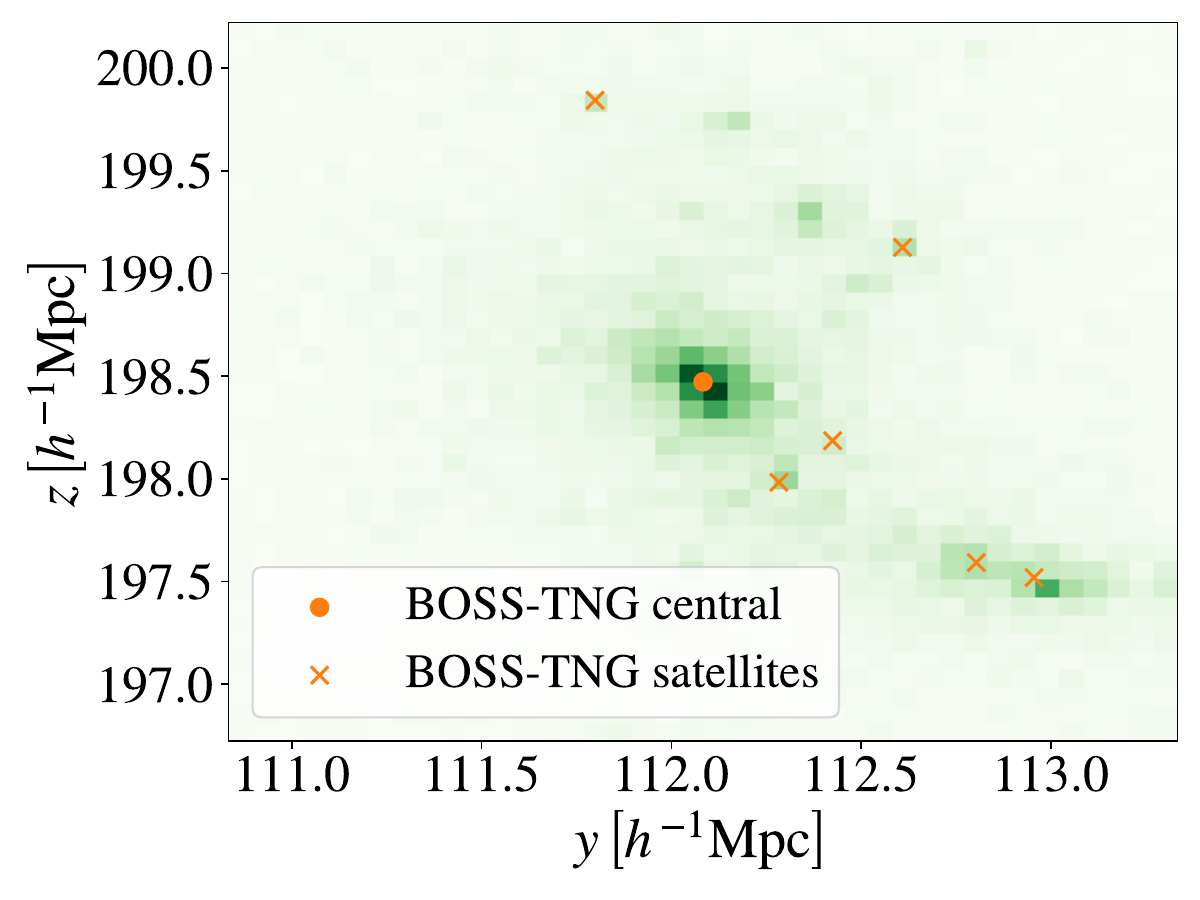}
    \caption{Distribution of \bosstng satellites and mass in a randomly selected halo with $\logMh=14.5$ mass. The green histogram indicates the mass distribution, with darker colours for increasing densities, whereas the dot and crosses indicate the position of central and satellite galaxies, respectively. We can readily see that \bosstng satellites are located at peaks of the density field and that their distribution is asymmetrical.
    }
    \label{fig:hod_slensing}
\end{figure}

Standard HOD models populate halos with satellites assuming that these either follow a spherically-symmetric profile or trace the mass distribution within halos in an unbiased fashion. However, $N$-body simulations show that the majority of dark matter halos are not symmetric \citep[e.g.,][]{jing2002TriaxialModelingHalo} and satellite galaxies sit at the centre of subhalos \citep[e.g.,][]{berlind2003HaloOccupationDistribution}, i.e., the relic of the satellite's host halo before accretion. In this section, we test these assumptions for \bosstng galaxies.

In Fig.~\ref{fig:hod_slensing}, we display the distribution of \bosstng satellites and mass in a randomly selected $\logMh=14.5$ halo. As expected, we find that the mass distribution is not symmetric and that \bosstng satellites are located at local peaks of the halo density field. The first effect causes HOD models to underestimate GGL because non-spherical halos in certain orientations are much better lenses than their spherical counterparts, a phenomenon known as halo triaxiality
\citep[e.g.,][]{oguri2005CanSteepMass, corless2007StatisticalStudyWeak}. Similarly, HOD models placing satellites outside subhalos underestimate GGL because doing so reduces the average density around satellites; this effect is commonly known as subhalo lensing \citep[e.g.,][]{zuMappingStellarContent2015}.


\subsection{Assembly bias}
\label{sec:hod_ab}

\begin{figure}
    \centering
    \includegraphics[width=\columnwidth]{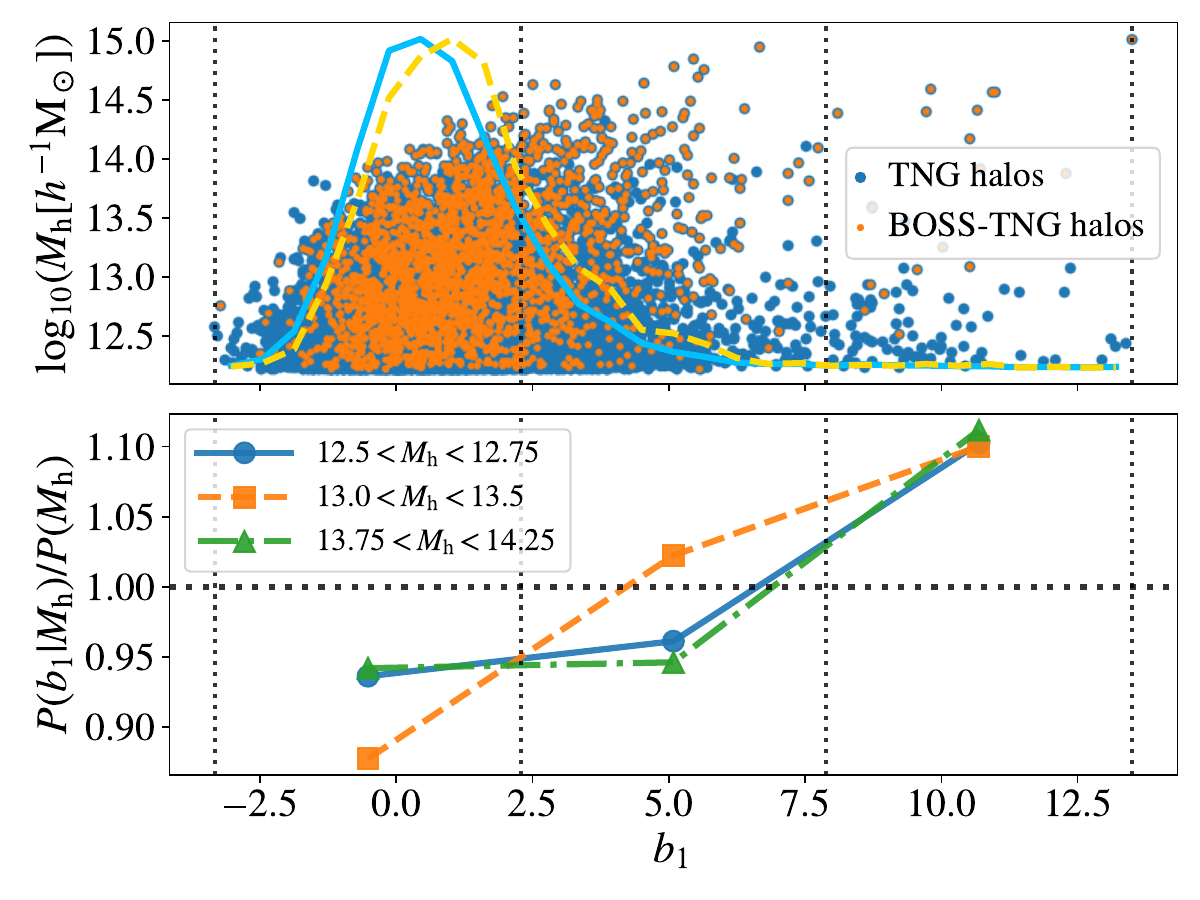}
    \caption{{\bf Top panel.} Impact of BOSS selection criteria on the $M_\mathrm{h}-b_1$ plane. Dots show the position of individual halos within this plane, while lines display the $b_1$ probability distribution function after marginalising over $M_\mathrm{h}$. The orange colour indicates the results for halos hosting \bosstng galaxies, while the blue colour does so for \tng halos with mass higher than the minimum mass of a halo hosting a \bosstng galaxy. {\bf Bottom panel.} Probability of finding a \bosstng galaxy in a halo with a particular large-scale bias and halo mass over the average probability for this mass. Each line indicates the results for a different bin in halo mass. As we can see, \bosstng galaxies preferentially reside in overdense environments for all halo masses.}
    \label{fig:hod_b1}
\end{figure}

\begin{figure*}
    \centering
    \includegraphics[width=\textwidth]{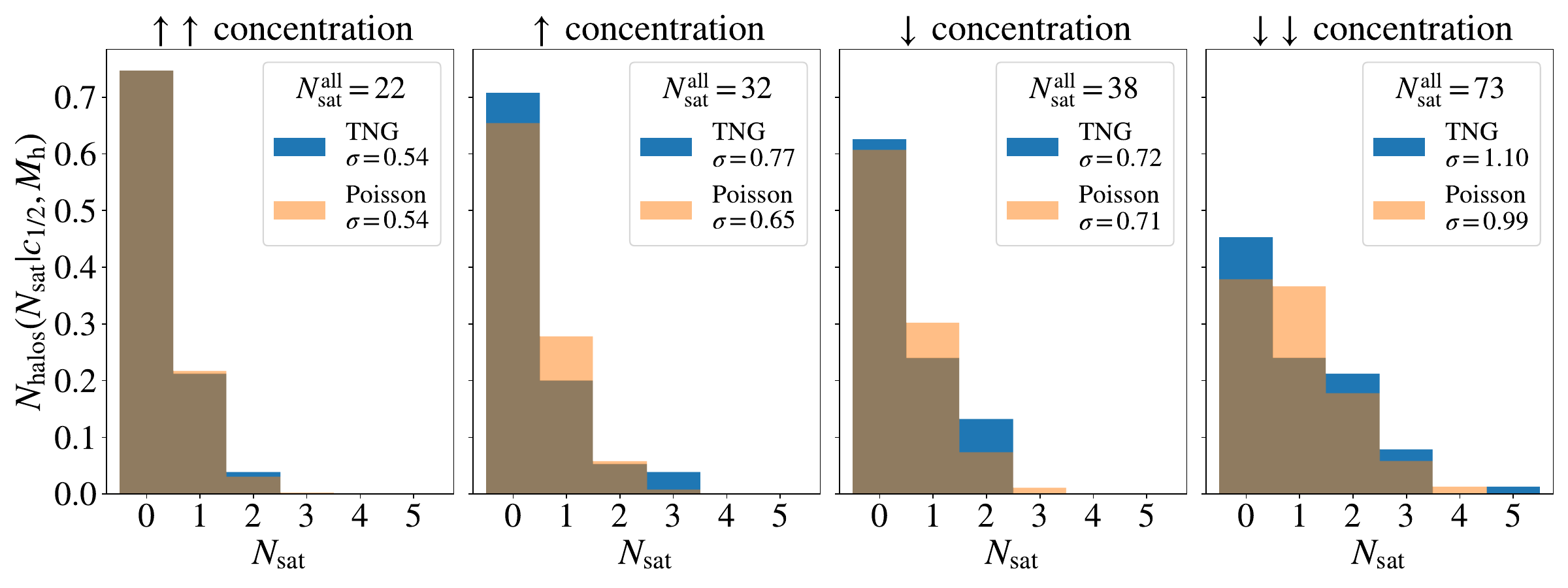}
    \caption{Number of \bosstng satellites hosted by halos within a narrow mass bin as a function of halo concentration. Each panel uses blue histograms to display the results for a different concentration quartile; we quote the total number of satellite galaxies in the legend of each panel. Orange histograms indicate predictions from a Poisson distribution. As we can see, \bosstng satellites preferentially populate halos with low concentration.}
    \label{fig:hod_conc}
\end{figure*}

It is by now firmly established that the large-scale clustering of haloes depends upon secondary halo properties in addition to halo mass \citep[halo assembly bias, e.g.,][]{shethEnvironmentalDependenceHalo2004, gao05, zheng05, wechsler06, gao07}. In turn, galaxy formation models predict that galaxies preferentially populate halos of a particular property at fixed halo mass, thereby propagating halo assembly bias into the galaxy population \citep[galaxy assembly bias, e.g.,][]{zhu06, croton07, zu08, hearin15, chavesmontero16, artale2018ImpactAssemblyBias}. However, standard HOD models ignore these effects and populate halos based solely on halo mass \citep[see, e.g.,][for more sophisticated HOD models] {hearin2016IntroducingDecoratedHODs, hadzhiyska2020LimitationsBasicHODb, hadzhiyska2022MillenniumTNGProjectImproveda, beltz-mohrmann2022AccurateModelingGalaxy}. In this section, we quantify the validity of this assumption for \bosstng galaxies.

To quantify assembly bias, we first compute the large-scale bias of each halo \citep{paranjape2018_HaloAssemblybias, contreras2021FlexibleModellingGalaxy}:
\begin{equation}
b_1({\bf x}) \equiv \left \langle \frac{V_\mathrm{box}}{P(|{\bf k}|)}  \, \exp(i\,{\bf k}\cdot {\bf x}) \delta^*({\bf k}) \right\rangle_{k < 0.2 h/{\rm Mpc}},
\end{equation}
where ${\bf x}$ is the location of a particular halo, $\delta^*$ is the complex conjugate of the matter density field in Fourier space, $P(k)$ the power spectrum of the matter density field as a function of wavenumber $k$, the angular brackets denote an average over scales smaller than $k=0.2 h/{\rm Mpc}$, $V_\mathrm{box}$ is the volume of the simulation box, and $i$ is the imaginary unit. We estimate the matter density field using the same diluted sample of mass particles as for computing GGL (see \S\ref{sec:methods_observables}).

By construction, the value of $b_1$ for a particular object is the density field at its position over that around a random location. Given that we only consider linear scales to compute it, the average value of $b_1$ for a galaxy sample equals the large-scale bias of this sample. The motivation for using this environmental property is that no internal halo property captures halo assembly bias completely \citep[e.g.,][]{gao07}.

In the top panel of Fig.~\ref{fig:hod_b1}, we display the impact of BOSS selection criteria on the $M_\mathrm{h}-b_1$ plane. Dots show the position of individual halos within this plane, while lines display the $b_1$ normalised probability distribution function after marginalising over $M_\mathrm{h}$. The orange colour indicates the results for halos hosting \bosstng galaxies, while the blue colour does so for \tng halos with a mass higher than the minimum mass of a halo hosting a \bosstng galaxy. As we can see, \bosstng galaxies preferentially populate halos with slightly larger $b_1$.

In the bottom panel of Fig.~\ref{fig:hod_b1}, we show the probability of finding a \bosstng galaxy in a halo with a particular $M_\mathrm{h}$ and $b_1$ over the average probability for this $M_\mathrm{h}$. Each line indicates the results for a different host halo mass bin. As we can see, \bosstng galaxies preferentially reside in overdense environments for all halo masses. Specifically, for halos more massive than $M_\mathrm{h}=13.4$, we find that the host halos of \bosstng galaxies present on average 9\% higher $b_1$ than \tng halos, and that this difference increases up to 32\% when we weigh \bosstng halos by the number of galaxies that these contain. Note that the probability of finding a \bosstng galaxy in halos more massive than $M_\mathrm{h}=13.4$ is greater than 50\%, and that the discrepancy in large-scale bias decreases for a lower mass cutoff.

Therefore, the strength of galaxy assembly bias for this sample is similar to that predicted by this and other galaxy formation models for stellar mass selected and quenched galaxies \citep[e.g.,][]{croton07, zentner14, chavesmontero16, contreras2019EvolutionAssemblyBias, contreras2021FlexibleModellingGalaxy, montero-dorta2021InfluenceHaloMass, hadzhiyska2021GalaxyAssemblyBiasa}. Nevertheless, observational constraints on galaxy assembly bias are not conclusive \citep[e.g.,][]{berlind2006PercolationGalaxyGroups, yang2006WeakLensingGalaxies, cooper2010GalaxyAssemblyBias, wang2013DetectionGalaxyAssembly, lacerna14, campbell2015AssessingColourdependentOccupation, hearin15, miyatake2016EvidenceHaloAssembly, saito16, tinker17, zu2017LevelClusterAssembly, busch2017AssemblyBiasSplashback, obuljen2020DetectionAnisotropicGalaxy, beltz-mohrmann2022AccurateModelingGalaxy}, and thus the extent of this effect may be different for BOSS galaxies.


\subsection{Occupational variation}
\label{sec:hod_conc}

In the previous section, we showed that \bosstng galaxies preferentially populate halos with large $b_1$. Here, we explore whether their occupancy depends upon internal halo properties at fixed halo mass \citep [see, e.g.,][for similar studies for other galaxy samples]{zehavi2018ImpactAssemblyBias, artale2018ImpactAssemblyBias, bose2019RevealingGalaxyhaloConnectiona, yuan2022IllustratingGalaxyhaloConnectiona}. Specifically, we study galaxy occupancy as a function of halo concentration because, at fixed halo mass, the GGL signal increases for galaxies in more concentrated halos \citep[e.g.,][]{xhakajMassDetectingSecondary2022}.

We estimate halo concentration using $c_{1/2}=r_{200}/r_{1/2}$, where $r_{1/2}$ refers to the halo half mass radius; note that this estimator is well-defined for unrelaxed halos. In Fig.~\ref{fig:hod_conc}, we display the number of \bosstng satellites in halos with masses within the interval $13.75<\logMh<14.25$. From left to right, panels show the results for quartiles in decreasing halo concentration; each quartile includes 75 halos. We find that halos in the least concentrated quartile contain 3.3 times more satellites than those in the most concentrated quartile, implying that standard HOD implementations overestimate the strength of GGL for the \bosstng sample.

We also find that the \bosstng satellite distribution departs from a Poisson distribution for increasingly less concentrated halos. We can visually see this by comparing the blue and orange histograms in each panel; the latter indicate the results for Poisson distributions with the same mean as the corresponding \bosstng distribution. For a more quantitative comparison, we quote the standard deviations of these distributions in the legend of each panel. Standard HOD models assume a Poissonian distribution for satellite galaxies (see in \S\ref{sec:methods_hod}); thus, the \bosstng sample presents halos with more satellites than HOD expectations. Because Poissonian deviations are only significant for low-concentration halos, this effect causes HOD models to further overestimate GGL for \bosstng galaxies.


\subsection{Baryonic effects}
\label{sec:hod_baryons}

\begin{figure}
    \centering    
    \includegraphics[width=\columnwidth]{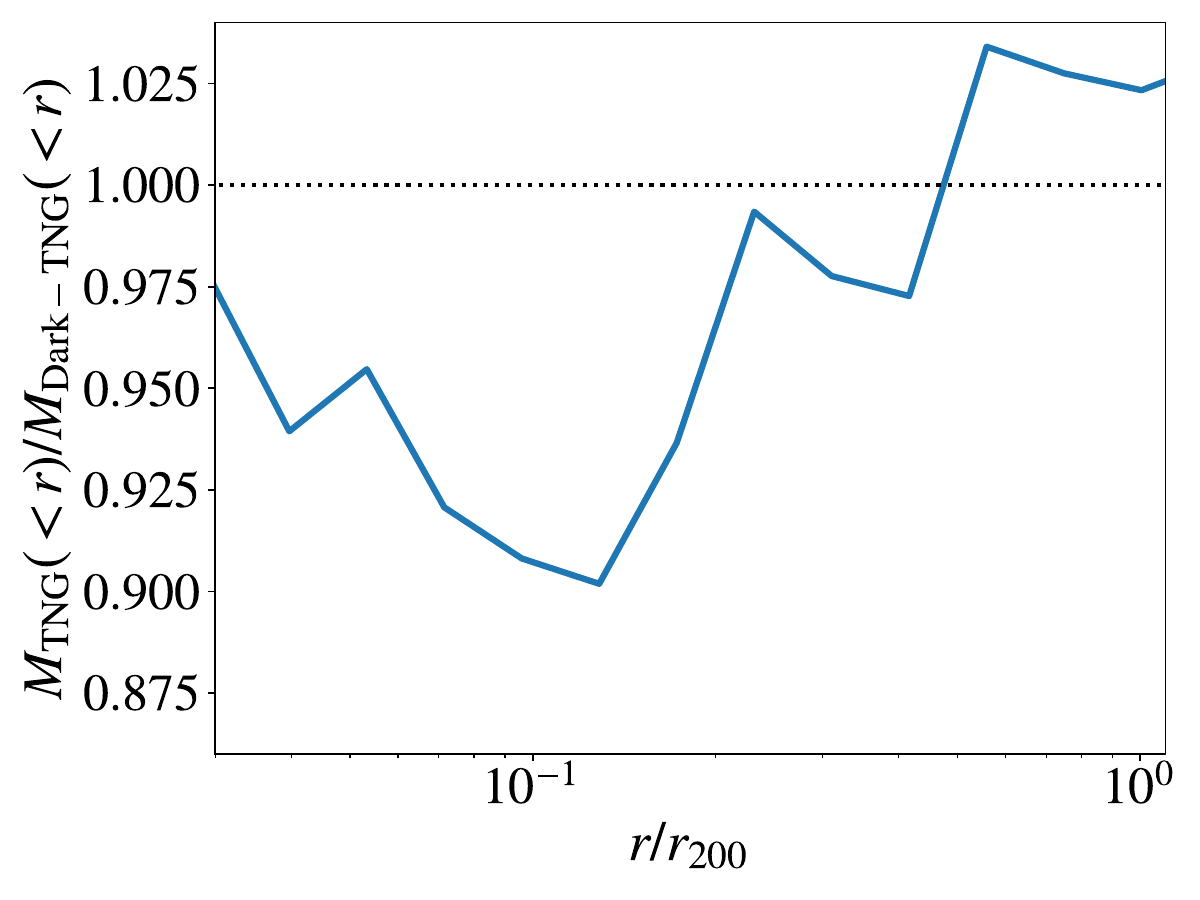}
    \caption{Average ratio between the cumulative mass distribution around the host halos of \bosstng galaxies in the \tng and \dtng simulations. As we can see, baryonic effects push mass towards the outer regions of halos, thereby decreasing the small-scale amplitude of GGL in \tng relative to \dtng.
    }
    \label{fig:hod_baryons}
\end{figure}

Baryonic effects decrease the clustering of matter on scales of the order of $1\Mpch$, thereby changing GGL on small scales \citep[e.g.,][]{chisari2019ModellingBaryonicFeedback}. However, most HOD analyses do not account for these effects. In this section, we quantify the magnitude of baryonic effects for \bosstng galaxies.

We isolate the impact of baryonic effects on the matter distribution by comparing the cumulative distribution of matter surrounding \bosstng galaxies in the \tng and \dtng simulations (see \S\ref{sec:methods_galaxies} for our approach to connecting galaxies between these simulations). We display the ratio between these distributions in Fig.~\ref{fig:hod_baryons}. As we can see, baryonic effects sweep mass from the inner to the outer regions of \bosstng halos, reducing the cumulative mass distribution by $\simeq10\%$ on scales of the order of $1\Mpch$. Consequently, we expect HOD models to overestimate GGL for \bosstng galaxies on small scales by about this percentage.


\section{Galaxy formation and \textbf{\textit{lensing is low}}}
\label{sec:lil}

In the previous section, we showed that HOD models fail to capture multiple galaxy formation effects predicted by the \tng simulation for BOSS-like galaxies. In this section, we quantify whether these shortcomings could originate the \lil problem. First, we create mock catalogues in \S\ref{sec:lil_catalogues} that we use to quantify the impact of different galaxy formation effects on GC, GGL and the \lil problem in \S\ref{sec:methods_hod}. Then, we study the origin of the \lil problem in \S\ref{sec:lil_origin}.


\subsection{Galaxy formation mocks}
\label{sec:lil_catalogues}

We create a mock galaxy catalogue to test each of the HOD shortcomings discussed in \S\ref{sec:hod}:

\begin{itemize}
    \item {\bf Parametric occupation.} To create this catalogue, we first measure $\langle N_\mathrm{cen}|\Mhalo \rangle$ and $\langle N_\mathrm{sat}|\Mhalo \rangle$ from the \bosstng catalogue by dividing \bosstng halos into 50 logarithmically-spaced halo mass bins with 0.06 dex width. Then, we populate \dtng halos with central and satellite galaxies using a nearest integer and a Poissonian distribution with the aforementioned means, respectively. Finally, we assign phase-space coordinates to mock galaxies in the same manner as our HOD implementation (see \S\ref{sec:methods_hod}). This catalogue mimics the mass-dependent occupation of \bosstng galaxies whereas satisfying any other assumption of our HOD implementation. Therefore, it serves to test errors arising from the limited precision of simple parametric forms describing the occupation distribution of \bosstng galaxies (see \S\ref{sec:hod_parametric}).

    \item {\bf Satellite segregation.} We create this catalogue by first measuring $\langle N_\mathrm{sat}|\Mhalo, r_\mathrm{cen}\rangle$ from \bosstng satellites. Then, we assign satellite galaxies to dark matter halo particles using this probability distribution; for central galaxies, we follow the same approach as our HOD implementation. This catalogue reproduces the mass-dependent occupation distribution of \bosstng central and satellite galaxies and the radial profile of the latter. Therefore, in addition to the flexibility of the HOD functional form, this catalogue tests the assumption that satellite galaxies follow the mass profile of their host halos (see \S\ref{sec:hod_sats}).

    \item {\bf Assembly bias.} To produce this catalogue, we first measure the average occupation distribution of \bosstng galaxies as a function of both mass and large-scale bias, i.e., $\langle N_\mathrm{cen}|\Mhalo, b_1 \rangle$ and $\langle N_\mathrm{sat}|\Mhalo, r_\mathrm{cen}, b_1 \rangle$. To do so, we use 10 linearly-spaced bins in $b_1$ with $\Delta b_1=1.9$ width and the same $\Mhalo$ binning as for the previous catalogues. Then, we randomly populate \dtng halos according to these probability distributions; therefore, this catalogue reproduces the level of galaxy assembly bias for \bosstng galaxies in addition to the properties captured by the previous catalogue. We use it to additionally test the impact of assuming no assembly bias for \bosstng galaxies.

    \item {\bf Occupational variation.} We create this catalogue by considering the role of halo concentration in addition to mass-dependent occupation, satellite segregation, and assembly bias. Analogously to previous cases, we measure $\langle N_\mathrm{cen}|\Mhalo, b_1, c_{1/2} \rangle$ for \bosstng central galaxies. To also capture deviations of the satellite occupation statistics from a Poissonian distribution (see \S\ref{sec:hod_conc}), we measure the probability of finding $N_\mathrm{sat}$ \bosstng satellite galaxies in each \tng halo,  $P(N_\mathrm{sat}|\Mhalo, r_\mathrm{cen}, b_1, c_{1/2})$. To do so, we use 10 linearly-spaced bins in $c_{1/2}$ with $\Delta c_{1/2}=0.1$ width, and the same binning in $\Mhalo$ and $b_1$ as for the previous catalogues. Finally, we randomly populate \dtng halos according to these probability distributions. In addition to the effects tested by the previous catalogues, this mock serves to quantify the impact of correlations between occupation number, satellite occupation statistics, and halo concentration.
    
    \item {\bf Baryonic effects.} This catalogue assigns to \bosstng galaxies the position of their counterparts in the \dtng simulation (see \S\ref{sec:methods_galaxies}). Therefore, we can isolate the impact of not considering baryonic effects in HOD models by comparing GC and GGL measurements from this catalogue and \bosstng galaxies. On the other hand, the strength of baryonic effects in the \tng simulation is weak compared to predictions from other simulations and observational estimates \citep[e.g.,][]{arico2020ModellingLargescaleMassa, amodeo2021AtacamaCosmologyTelescope, Chen:2022}. To test the range of baryonic effects compatible with observations, we also generate two catalogues that reproduce baryonic effects as predicted by the low- and high-AGN simulations of the BAHAMAS suite. We do so by modifying the mass distribution in the \dtng simulation (see \S\ref{sec:methods_baryons}).

    \item {\bf Subhalo lensing and halo triaxiality.} To estimate the impact of these effects on GC and GGL, we randomly perturb the three-dimensional positions of the \dtng counterparts of \bosstng satellites while holding fixed the distance to the centre of their host halos. As a result, satellite galaxies leave subhalo potential wells, and their distribution becomes spherically symmetric. Therefore, this catalogue tests the joint impact of not incorporating baryonic effects, subhalo lensing, and halo triaxiality in HOD models.
\end{itemize}

We compute the projected correlation function and excess surface density for the previous catalogues following \S\ref{sec:methods_observables}. Then, for all mock catalogues except the one targeting baryonic effects, we repeat this procedure 300 times for GC and 30 times for GGL using different random draws to reduce stochastic noise. Note that we use the matter density field as measured in the \dtng simulation for GGL measurements.


\subsection{Influence of galaxy formation on GC and GGL}
\label{sec:lil_impact}

\begin{figure}
    \centering
    \includegraphics[width=\columnwidth]{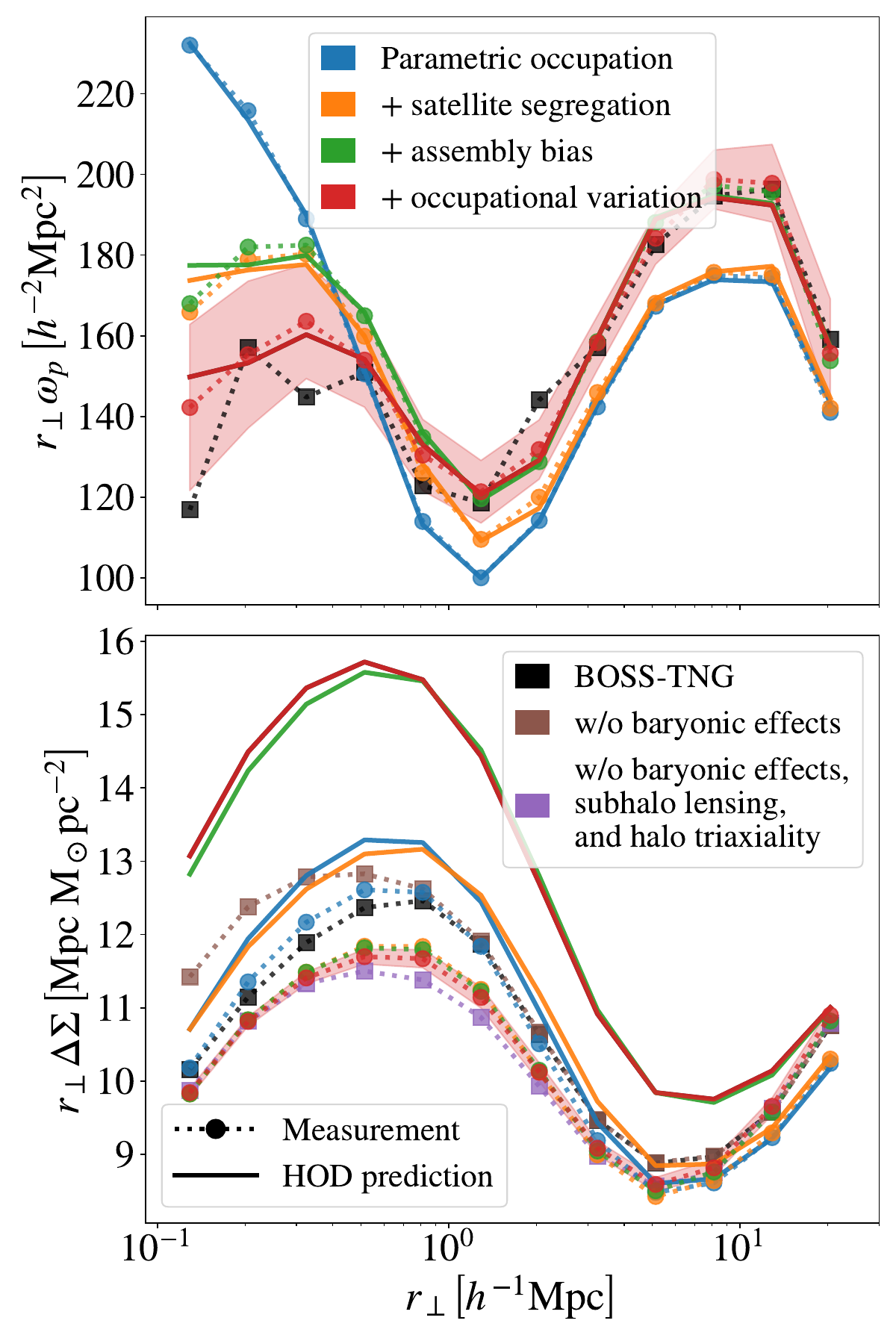}    \includegraphics[width=\columnwidth]{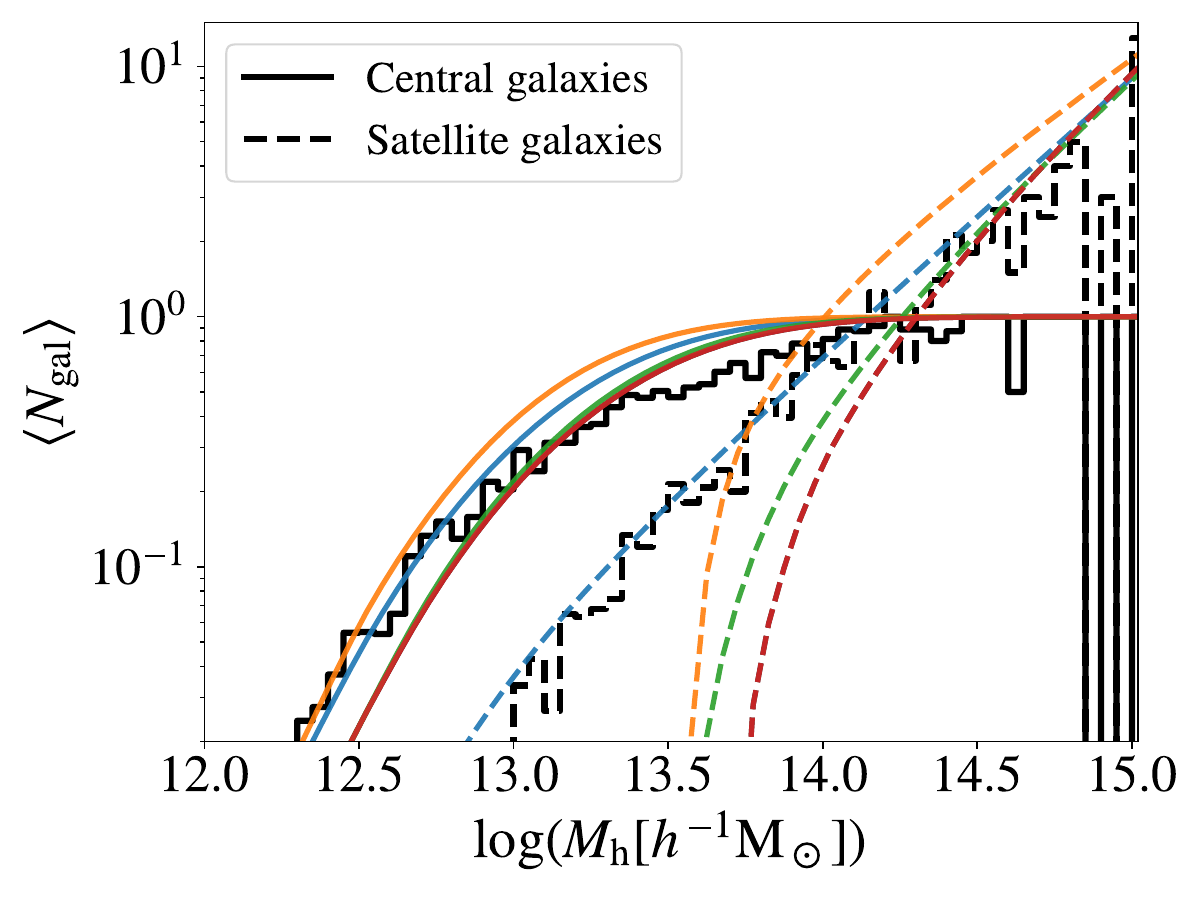}
    \caption{Influence of galaxy formation effects not modelled by standard HOD implementations on GC (top panel), GGL (middle panel), and halo occupation distribution (bottom panel). The black colour indicates the results for \bosstng galaxies, whereas other colours do so for mock catalogues testing different HOD assumptions (see legend). Dotted lines show simulation measurements, solid lines display predictions from best-fitting HOD models to GC, and shaded areas denote $1\sigma$ uncertainties for one of the mock catalogues. In the bottom panel, solid and dashed lines indicate the occupation distribution of central and satellite galaxies, respectively. Our standard HOD implementation is flexible enough to reproduce the clustering of all mock catalogues precisely; however, it increasingly overpredicts GGL as more galaxy formation effects are considered.
    }
    \label{fig:wp_ds_hod}
\end{figure}

\begin{figure*}
    \centering \includegraphics[width=0.8 \textwidth]{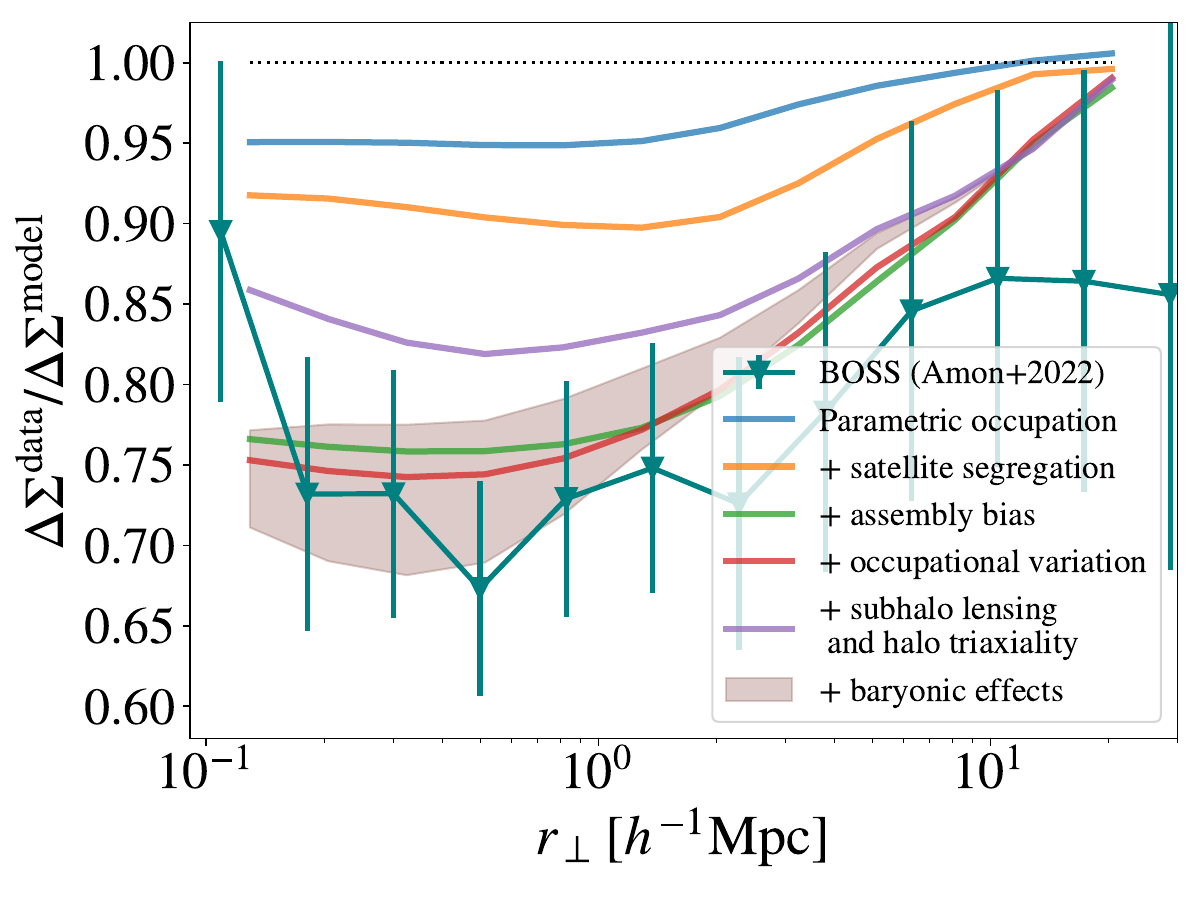}
    \caption{Ratio between GGL measurements from colour-selected galaxies and predictions from best-fitting HOD models to their clustering. The dark green line displays the results for BOSS galaxies \citepalias{amon2022ConsistentLensingClustering}, whereas other lines do so for \bosstng galaxies when considering an increasing number of galaxy formation effects. Error bars indicate $1\sigma$ uncertainties; the shaded area delimits the range of baryonic effects compatible with observations. We find that standard HOD models are not complex enough to describe the galaxy-halo connection for \bosstng galaxies precisely, which generates a \lil problem compatible with that found for BOSS galaxies.}
    \label{fig:lil_ds}
\end{figure*}

In what follows, we use the mock catalogues created in the previous section to test the impact of standard assumptions of HOD models on GC, GGL, and the \lil problem.

In the top panel of Fig.~\ref{fig:wp_ds_hod}, we use dotted lines to display GC measurements from the \bosstng sample (black line) and mock catalogues (coloured lines). As we can see, galaxy formation effects modify GC significantly, changing the amplitude of $\wps$ up to $\simeq60\%$. On small scales, satellite segregation and occupational variation reduce $\wps$ by $\simeq40$ and 20\%, respectively; this is the result of \bosstng galaxies preferentially populating the outer regions (see \S\ref{sec:hod_sats}) of less concentrated halos (see \S\ref{sec:hod_conc}). On the other hand, assembly bias increases galaxy clustering by approximately 10\% on large scales because \bosstng galaxies preferentially populate halos in overdense regions (see \S\ref{sec:hod_ab}). Notably, the clustering of the mock sample capturing the mass-dependent occupation, satellite distribution, assembly bias, and concentration-dependent occupation of \bosstng galaxies (red line) agrees with the clustering of \bosstng galaxies within the red shaded area, which denotes the standard deviation of GC measurements from 300 versions of this mock using different random draws (see \S\ref{sec:lil_catalogues}). We thus conclude that these four effects govern the clustering of colour-selected galaxies in the \tng simulation. We checked that the impact of baryonic effects, subhalo lensing, and halo triaxiality on GC is negligible.

In the middle panel of Fig.~\ref{fig:wp_ds_hod}, we show that assembly bias increases GGL on large scales whereas satellite segregation, baryonic effects, subhalo lensing, and halo triaxiality modify it on small scales. Satellite segregation decreases the signal because satellite galaxies preferentially sit on the outskirts of halos where projected densities are lower than in the centre, reducing each satellite's contribution (see \S\ref{sec:hod_sats}). Similarly, baryonic effects decrease GGL because these reduce the amount of mass on the inner regions of halos (see \S\ref{sec:hod_ab}). On the other hand, subhalo lensing and halo triaxiality increase the signal because satellite galaxies sit on local peaks of the density field and halos in certain orientations are better lenses than their spherical counterparts (see \S\ref{sec:hod_slens}). Remarkably, GGL measurements from \bosstng galaxies in the absence of baryonic effects, subhalo lensing, and halo triaxiality agree with those from the mock reproducing the mass-dependent occupation, satellite distribution, assembly bias, and concentration-dependent occupation of \bosstng galaxies within uncertainties. Thus, the previous effects are enough to reproduce GGL in the \tng simulation.

Despite the strong impact of some of these effects on GC, we can readily see that our HOD implementation reproduces the clustering of all mock catalogues accurately\footnote{We do not display uncertainties in HOD predictions because we use a covariance matrix ensuring robust inference but not capturing the level of uncertainties in any particular survey (see \S\ref{sec:methods_emulator}).}. In the bottom panel of this figure, we show that the success of HOD models comes at the cost of not describing the occupation distribution of \bosstng galaxies accurately. Black histograms display their occupation distribution, whereas coloured lines depict predictions from best-fitting HOD models to GC. As we can see, these models underpredict the average central (satellite) occupation of low-mass halos to accommodate the increase in large-scale (decrease in small-scale) clustering caused by assembly bias (satellite segregation and concentration-based occupation).

Interestingly, best-fitting HOD models overpredict the GGL signal for mock catalogues. This failure results from inaccuracies in the occupation distribution predicted by best-fitting models: these populate low-mass halos with too few satellites, which increases the average halo mass of the sample and, consequently, the GGL signal. Concurrently, all galaxy formation effects considered except subhalo lensing and halo triaxiality decrease the GGL signal measured from the catalogues, further exacerbating the gap between mock measurements and HOD predictions. We therefore conclude that the combination of HOD inaccuracies and galaxy formation effects creates a \lil problem. In the following section, we investigate whether the magnitude of this tension is consistent with observational estimates.


\subsection{The origin of the \textbf{\textit{lensing is low}} problem}
\label{sec:lil_origin}


\begin{figure}
    \centering 
    \includegraphics[width=\columnwidth]{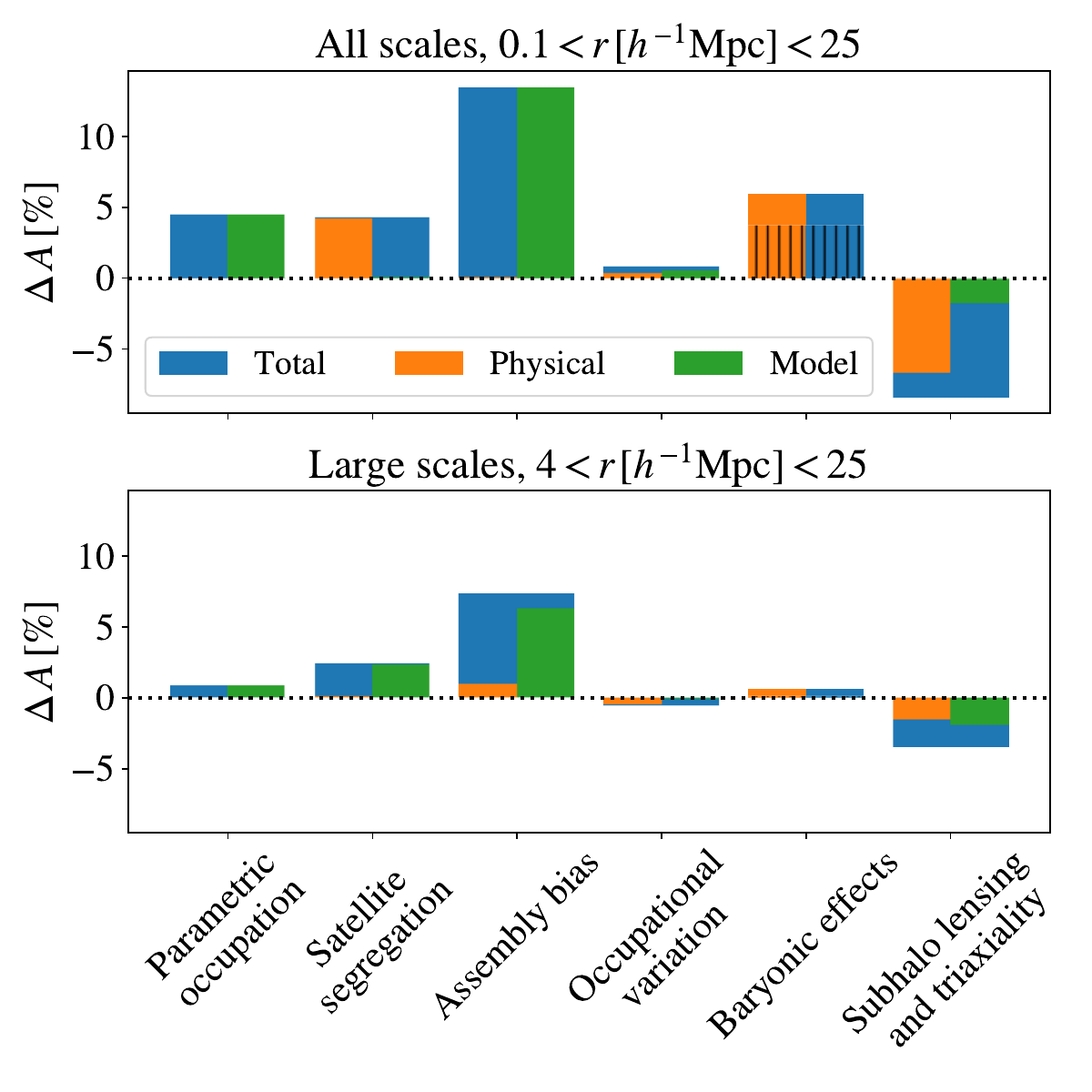}
     \caption{Impact of galaxy formation effects on the \lil problem when considering all scales (top panel) and only large scales (bottom panel). Blue bars indicate the total tension caused by each effect, whereas orange and green bars display the fraction due to variations in GGL measurements and HOD predictions, respectively. As we can see, assembly bias is the most important effect driving the \lil problem, followed by satellite segregation and baryonic effects.}
    \label{fig:lil_bars}
\end{figure}

In Fig.~\ref{fig:lil_ds}, we display the ratio between GGL measurements around observational and mock galaxies and predictions from best-fitting HOD models to their GC. The line with error bars displays the results for BOSS galaxies \citepalias{amon2022ConsistentLensingClustering}, whereas other lines do so for \bosstng galaxies when using a standard HOD implementation that fails to incrementally capture their mass-dependent occupation, satellite distribution, assembly bias, concentration-dependent occupation, subhalo lensing, halo triaxiality, and baryonic effects. Remarkably, we find that HOD inaccuracies caused by these galaxy formation effects produce a \lil problem that is fully consistent with that found in recent observational studies. Therefore, the \tng simulation predicts that the \lil problem is primarily (if not entirely) caused by the oversimplistic nature of standard HOD models. We emphasise that this prediction is based on our current best understanding of galaxy formation, as the \tng simulation is one of the most advanced galaxy formation models.

As in \S\ref{sec:btng_lil}, we further quantify the \lil problem by computing the weighted average of the previous ratio. We find that galaxy formation effects induce a $A=0.73$-0.80 tension depending on the strength of baryonic effects, which agrees with recent observational constraints: $A=0.76\pm0.03$ \citepalias{amon2022ConsistentLensingClustering}. This close agreement remains when only scales larger than $4\,\Mpch$ are considered: we find $A=0.91$ for \bosstng galaxies and $0.86\pm0.06$ for \citetalias{amon2022ConsistentLensingClustering}. Assembly bias is the only galaxy formation effect affecting the 2-halo term, and thus we naturally expect a decrease in the tension when only considering scales larger than $4\,\Mpch$.

In Fig.~\ref{fig:lil_bars}, we summarise the contribution of each galaxy formation effect to the \lil problem. Blue bars display the total contribution of each effect, whereas the orange bars show the fraction from physical variations -- how much this effect changes GGL -- and green bars from model inaccuracy -- the systematic error caused by the best-fitting HOD model preferring an inaccurate occupation distribution. The top and bottom panels display the results when considering all scales and only those larger than $4\,\Mpch$, respectively. We can readily see that assembly bias is the most important effect: it generates a tension of $\Delta A=0.14$ on all scales and $\Delta A=0.07$ on just large scales. On small scales, baryonic effects generate slightly more tension than parametric occupation and satellite segregation, while subhalo lensing and halo triaxiality alleviate it by approximately $\Delta A=0.10$.

We proceed to use these findings to understand why the \lil problem presents no redshift evolution from $z=0.7$ to 0.2 \citepalias[e.g.,][]{lange2019_NewPerspectivesBOSS, amon2022ConsistentLensingClustering}. To do so, we focus on the two effects producing most of the tension: assembly bias and baryonic effects. On the one hand, the strength of galaxy assembly bias presents little evolution for BOSS-like galaxies from $z=1$ to 0 \citep[see fig.~5 of][]{montero-dorta2021InfluenceHaloMass}. On the other hand, the cosmological hydrodynamical simulations \tng and BAHAMAS predict little redshift evolution within this redshift range \citep[see fig.~7 of][]{arico2021SimultaneousModellingMatter}. Taken together, these results explain the lack of significant redshift evolution in the strength of the \lil problem.

Throughout this section, we studied multiple galaxy formation effects affecting the galaxy-halo connection of \bosstng galaxies. However, some effects like miscentering are not present in the \tng simulation as, by construction, central galaxies have the coordinates of the centre of potential of their host halos. Furthermore, some HOD models consider different assumptions relative to our implementation; for example, the \citetalias{amon2022ConsistentLensingClustering} model populates halos with satellites only if these halos contain a central. In Appendix~\ref{app:small}, we study the influence of this assumption and miscentering on the \lil problem.


\section{Summary and conclusions}
\label{sec:conclusions}

We investigated the consistency between galaxy-galaxy lensing (GGL) measurements from colour-selected galaxies and GGL predictions from galaxy-halo connection models optimised to describe their clustering. Our aim was to shed light on the so-called \lil tension, in which GGL measurements around BOSS galaxies are from 20 to $40\%$ lower than theoretical expectations.

To explore this issue, we first selected a sample of galaxies from the cosmological hydrodynamical simulation IllustrisTNG that mimicked the properties of BOSS galaxies (\bosstng sample). Then, we performed GGL measurements on \bosstng galaxies and compared these to predictions from best-fitting Halo Occupation Distribution (HOD) models to their projected clustering. Interestingly, we found that \bosstng galaxies exhibited a \lil problem with similar magnitude and scale dependence as that found for BOSS galaxies (see Fig.~\ref{fig:boss_lil}). Throughout the remainder of this work, we focused on understanding the origin of this problem.

First, we tested the validity of standard HOD assumptions for the galaxy-halo connection of \bosstng galaxies. We found that
\begin{itemize}
    \item Standard HOD parametric forms do not capture the occupation distribution of \bosstng central galaxies precisely (parametric occupation; see Fig.~\ref{fig:hod_parametric}).

    \item The radial profile of \bosstng satellites is flatter than that typically assumed in HOD models (satellite segregation; see Fig.~\ref{fig:hod_radial}). 

    \item \bosstng satellites sit at local peaks of the density field in non-spherical halos, whereas HOD models assume that these galaxies are randomly distributed within spherically-symmetric halos (subhalo lensing and halo triaxiality; see Fig.~\ref{fig:hod_slensing}).

    \item At a fixed halo mass, \bosstng galaxies are preferentially hosted by halos in overdense regions (assembly bias; see Fig.~\ref{fig:hod_b1}) and low concentration (occupational variation; see Fig.~\ref{fig:hod_conc}), whereas standard HOD models assume no occupational dependence on these properties.

    \item The host halos of \bosstng galaxies present less mass in their inner regions than HOD expectations (baryonic effects; see Fig.~\ref{fig:hod_baryons}).
\end{itemize}

We found that these galaxy formation effects modify GC significantly: satellite segregation and occupational variation reduce small-scale clustering by up to $\simeq60\%$, whereas assembly bias increases large-scale clustering by about 10\%. Despite these variations, standard HOD implementations are sufficiently flexible to capture the projected clustering of \bosstng galaxies across the entire range of scales we considered ($0.1 < r [\hMpc] < 25$). To accomplish this, best-fitting HOD models assume an incorrect galaxy occupation distribution, causing these models to overestimate GGL measurements (see Fig.~\ref{fig:wp_ds_hod}). Therefore, HOD inaccuracies resulting from incorrect modelling of the previous galaxy formation effects generate a \lil problem. Notably, the extent of this problem is in remarkable agreement with observational estimates (see Fig.~\ref{fig:lil_ds}). We emphasise that these effects are predicted by a fully consistent $\Lambda$CDM hydrodynamical simulation and thus by our best current understanding of how galaxies form and populate haloes.

However, the \tng simulation was not optimised to match all details of BOSS galaxies, and thus there may be differences between \tng predictions and the actual influence of galaxy formation effects on the \lil tension. Independently of this, the previous effects are generic features of galaxy formation models and therefore must be considered for precise consistency relations between GC and GGL. Of course, this requirement is essential when extracting unbiased cosmological constraints from small scales and generating mock catalogues to validate survey analysis pipelines.


\section*{Acknowledgements}

We thank David Alonso, Antonio Montero-Dorta, Shun Saito, and Simon White for their useful comments and discussion. We thank Johannes Lange and Naomi Robertson for sharing their data. We acknowledge the work of Giovanni Aric\`o in developing the baryonification module of the BACCO package and his assistance when running it. We also thank the SDSS survey and the IllustrisTNG collaboration for making their data publicly available. We gratefully acknowledge the use of the Atlas EDR cluster at the Donostia International Physics Center (DIPC). JCM and REA acknowledge support from the ERC-StG number 716151 (BACCO). JCM acknowledges support from the European Union’s Horizon Europe research and innovation programme (COSMO-LYA, grant agreement 101044612), REA from the Project of excellence Prometeo/2020/085 from the Conselleria d'Innovaci\'o, Universitats, Ci\`encia i Societat Digital de la Generalitat Valenciana, and SC from ``Juan de la Cierva Incorporac\'ion'' fellowship (IJC2020-045705-I). IFAE is partially funded by the CERCA program of the Generalitat de Catalunya.


\section*{Data availability}

Our datasets will be shared upon request to the corresponding author.


\section*{Software}

This work made direct use of the following software packages: {\sc corrfunc} \citep{sinha2020_CORRFUNCSuiteblazing}, {\sc emcee} \citep{foremanmackey13}, {\sc ipython} \citep{perez2007_IPythonSystemInteractive}, {\sc matplotlib} \citep{hunter2007_Matplotlib2DGraphics}, {\sc mpi4py} \citep{dalcin2005_MPIPython, dalcin2008_MPIPythonPerformance, dalcin2011_ParallelDistributedcomputing, dalcin2021_Mpi4pyStatusUpdate}, {\sc numpy} \citep{harris2020_ArrayProgrammingNumPy}, {\sc pydoe2} \citep{rickardsjogrenanddanielsvensson2018PyDOE2ExperimentalDesign}, \textsc{pytorch} \citep{pytorch}, and {\sc scipy} \citep{virtanen2020_SciPyFundamentalalgorithms}.


\bibliographystyle{mnras}
\bibliography{biblio}

\appendix
\renewcommand{\thefigure}{A\arabic{figure}}

\begin{figure}
    \centering    
    \includegraphics[width=\columnwidth]{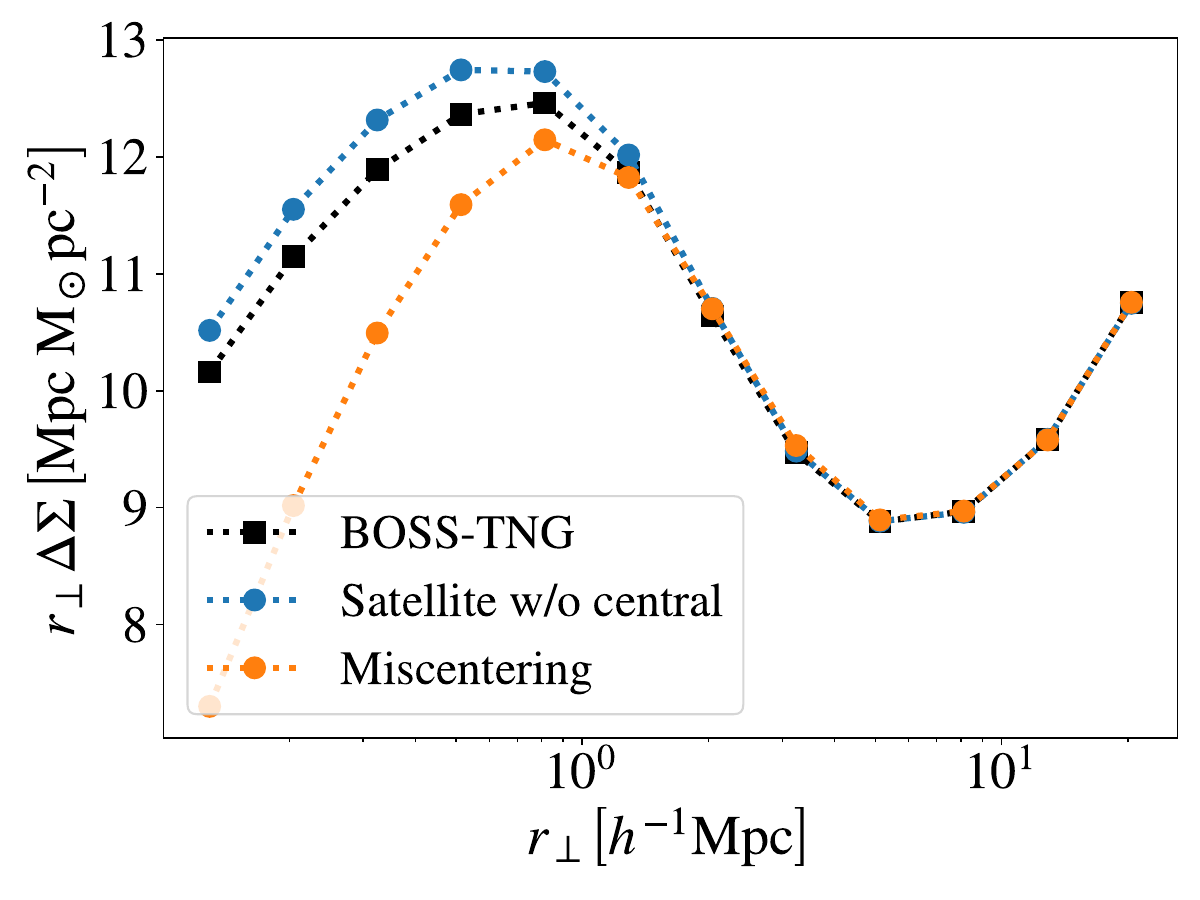}
    \caption{Impact of miscentering-related HOD assumptions on the \lil problem. The black line shows the GGL signal for \bosstng galaxies, the blue line does so for \bosstng galaxies after displacing satellite galaxies alone in their halos to the centre of these, and the orange line displays the GGL signal for \bosstng galaxies after moving central galaxies from the centre of potential to the centre of mass of their halos. We find that both effects contribute to the \lil problem on small scales.
    }
    \label{fig:hod_miscentering}
\end{figure}

\section{Other HOD assumptions}
\label{app:small}

In this section, we study whether two assumptions of some HOD implementations contribute to the \lil problem: satellite galaxies must reside in halos with a central galaxy, and central galaxies sit at the centre of their dark matter halos.

Some \lil studies assume that satellite galaxies only populate halos hosting a central galaxy \citepalias[e.g.,][]{amon2022ConsistentLensingClustering}; nevertheless, a significant fraction of \bosstng satellites reside in halos with no central (see \S\ref{sec:hod_parametric}). To check the impact of this assumption on the \lil problem, we create a \bosstng-based mock in which we assign the coordinates of their host halos to satellites in halos with no central. In Fig.~\ref{fig:hod_miscentering}, the blue and black lines display the GGL signal for this sample and \bosstng galaxies, respectively. As we can see, assuming that satellites only populate halos hosting a central may overestimate the GGL of \bosstng galaxies up to $3.5\%$ on scales below $1\Mpch$. We find that this assumption does not modify GC, and thus it contributes to the \lil problem on small scales.

Most HOD models assume that central galaxies sit at the bottom of the potential well of their host halos; nonetheless, some central galaxies are slightly displaced relative to the halo potential minimum \citep[e.g.,][]{yang2006WeakLensingGalaxies, johnston2007CrossCorrelationLensingDetermining, hilbert2010AbundancesMassesWeaklensing, saro2015ConstraintsRichnessmassRelation, zhang2019DarkEnergySurveyed}. For a broad estimate of the impact of miscentering on the \lil problem, we create a \bosstng-based mock in which we displace central galaxies from the centre of potential to the centre of mass of their host halos. We find that miscentering modifies GGL as much as $25\%$ on scales below $1\Mpch$, whereas it leaves GC unchanged. Consequently, HOD models not accounting for this effect will overestimate GGL on small scales, which exacerbates the \lil problem.

\bsp
\label{lastpage}
\end{document}